%% file: main_revised.tex
\documentclass[10pt,journal]{IEEEtran}
\usepackage[T1]{fontenc}
\usepackage[breaklinks,colorlinks,
            citecolor=blue,
            hypertexnames=false]{hyperref}
\usepackage[numbers]{natbib}
\usepackage{amsmath,amssymb,amsfonts}
\usepackage{bbm}
\usepackage{algorithm}
\usepackage{algorithmic}
\usepackage{array}
\usepackage{multirow}
\usepackage{booktabs}
\usepackage{longtable}
\usepackage{tikz}
\usepackage{tabularx}
\usetikzlibrary{positioning,arrows.meta,fit,backgrounds,calc}
\usepackage{tablefootnote}
\usepackage{colortbl}
\usepackage{graphicx}
\usepackage{adjustbox}
\usepackage{textcomp}
\usepackage{stfloats}
\usepackage{url}
\usepackage{verbatim}
\usepackage{xcolor}
\usepackage{xspace}
\usepackage{inconsolata}
\usepackage{tcolorbox}
\usepackage{threeparttable}
\usepackage{lmodern}
\usepackage{paralist}
\usepackage{multicol}
\usepackage{pifont}
\usepackage{subcaption}

\definecolor{mygray}{gray}{0.9}
\definecolor{gatecolor}{RGB}{216,90,48}
\definecolor{corecolor}{RGB}{83,74,183}
\newcolumntype{L}{>{\raggedright\arraybackslash}X}
\newcolumntype{P}[1]{>{\raggedright\arraybackslash}p{#1}}
\newcolumntype{C}[1]{>{\centering\arraybackslash}p{#1}}
\DeclareRobustCommand{\cmark}{\ding{51}}
\DeclareRobustCommand{\xmark}{\textcolor{black!35}{\ding{55}}}
\DeclareRobustCommand{\pmark}{\tikz[baseline=-0.55ex]{\draw[line width=0.5pt] (0,0) circle (0.8ex); \fill (0,0.8ex) arc (90:270:0.8ex) -- cycle;}}

\begin{document}
\input{preamble}

\title{Great Listeners, Poor Decision-Makers:\\ Rethinking Agentic Audio AI}

\author{Siddique Latif, Moid Sandu, Farrukh Pervez, Raja Jurdak}

\markboth{under-review}%
{Great Listeners, Poor Decision-Makers: Rethinking Agentic Audio AI}

\maketitle

\begin{abstract}
Audio artificial intelligence (AI) now recognizes speech, sounds, and music remarkably well, yet voice agents still act when they should not: they execute commands addressed to others, act on misheard content, and commit before the user finishes speaking. This position article argues that the bottleneck in agentic audio AI has shifted from recognition to decision-making. We propose a reference architecture with an explicit decision layer, separate from perception and reasoning, that chooses whether to act, wait, ask, or remain silent from evidence on addressedness, speaker identity, content confidence, tool reliability, utterance completeness, and action risk. We motivate the design through six recurring decision failures and a decision-theoretic argument that transcript confidence alone is insufficient to justify an action. We implement the layer with fuzzy inference, evolutionary optimization, and sequential decision-making, and evaluate it through simulations, ablations, and recorded audio, including egocentric recordings. Addressedness and speaker evidence matter most, risk-aware gating suppresses costly errors, and well-tuned policies perform similarly, with fuzzy gates more robust to overconfident addressedness estimates. On recorded audio, explicit gating cuts executed side talk from 41\% to 7--10\% while serving 83--85\% of valid requests under successful clarification, outperforming prompting-based baselines. These results support explicit action gating as an architectural requirement for reliable voice agents.
\end{abstract}


\IEEEPARstart{S}{uppose} a user asks a voice assistant to send a package to fifteen Oak Street. Background noise blurs the number, the assistant hears ``fifty,'' and the order goes to the wrong address \cite{ivry2026voice}. No component malfunctioned: the recognizer produced a plausible word, the language model planned a sensible action, and the tool executed correctly. What was missing was a step that asked whether the evidence justified an action that was costly and hard to undo.

This scenario captures our central argument. Audio artificial intelligence (AI) has advanced rapidly in \emph{listening}: large audio-language models (LALMs) now recognize speech, sounds, and music within a single model \cite{tang2023salmonn, chu2024qwen2}, and audio agents couple them with planning and tool use \cite{kuan2024speechcopilot, wijngaard2025audiotoolagent, wang2026audiomind, wang2026hiu}.

Decision-making has not kept pace. On 278 grounded customer-service tasks, a text-only reasoning agent succeeds on 85\% of tasks, whereas full-duplex voice agents succeed on 31--51\% with clean audio and 26--38\% under realistic acoustics, with 79--90\% of failures attributed to agent behavior \cite{ray2026tauvoice}. Studies that isolate the decision are starker. An audio language model that selects the correct tool in 96--97\% of control cases remains silent in only 1\% of cases when the same command comes from a different speaker \cite{zhang2026vgbench}: it understands the words but ignores the acoustic context that should govern whether to act. On egocentric smart-glasses recordings, real-time speech language models reach only 29--59\% accuracy, and in the benchmark authors' own model, multi-microphone spatial cues raise side-talk rejection from 85.4\% to 93.9\% \cite{lin2026wearvox}. The evidence needed to decide thus lies in the signal, not only in the words. Yet current LALMs underuse it: when lexical and acoustic cues conflict, they follow the words and neglect how they were spoken \cite{chen2026listen}.

Existing literature has not addressed this gap for audio. Surveys of agentic AI cover autonomous decision-making, tool use, planning, and multi-agent coordination, both in general \cite{abou2025agentic, schneider2025generative, arunkumar2026agentic} and in domains such as healthcare \cite{xu2025comprehensive}, but focus on text-based agents \cite{biswas2025building}. Recent work studies when text agents should abstain \cite{liu2026agentabstain, luo2026agenticabstention}, but there the uncertainty lies in the instruction, not in the signal and its source. Spoken dialogue systems have long confirmed uncertain requests \cite{bohus2005sorry, williams2007pomdp}, but they assume a single user in a turn-based exchange, not open, multi-party audio acted on through tools (Section~\ref{sec:sds}). Closest to ours, Ivry et al.~\cite{ivry2026voice} trace how acoustic stress propagates from signal degradation to interaction and action, and propose paired stressed and clean evaluation. We build on their diagnosis but ask a complementary question: not how degradation causes failures, but how an agent should \emph{decide} under the uncertainty that remains. This question leads to a formal decision model, an explicit decision layer, and computational intelligence (CI) methods to implement it.

Our claim is comparative. Audio agents are not flawless listeners, as these results show, but recognizing \emph{what} was said has advanced faster than deciding \emph{what to do} about it, and the latter now limits performance: an agent that transcribes perfectly still fails if it acts on words meant for someone else. Because the field moves quickly, several cited sources are 2026 preprints; we rest our central claims on refereed work \cite{ray2026tauvoice, lin2026wearvox, chen2026listen, liu2026agentabstain} and treat preprint results, including the VGBench benchmark used in part of our experiments \cite{zhang2026vgbench}, as preliminary.

\begin{tcolorbox}[colback=mygray,colframe=black!60,boxrule=0.4pt,arc=1pt,left=4pt,right=4pt,top=3pt,bottom=3pt]
\textbf{Position.} Audio agents have become far better listeners than decision-makers. The next advance in agentic audio AI will come less from better perception than from an explicit, uncertainty-aware decision layer that determines \emph{whether}, \emph{when}, and \emph{how} an agent acts on what it hears.
\end{tcolorbox}

This is a natural problem for CI. Decisions in audio agents must fuse graded and conflicting evidence, such as partial transcript confidence, uncertain speaker identity, and disagreement between tools, while trading off accuracy, latency, user effort, and risk. Fuzzy systems were designed for decisions under graded uncertainty \cite{zadeh1965fuzzy, bellman1970decision}, evolutionary computation for multi-objective design \cite{deb2002nsga2}, and reinforcement learning for sequential decisions with delayed consequences \cite{sutton2018rl}. Section~\ref{sec:ci} shows how each implements part of the decision layer.

Centered on a reference architecture with an explicit decision layer, this article makes four contributions:
\begin{enumerate}
\item \textbf{A diagnosis} of six recurring decision failures in current audio agents, with a decision-theoretic argument that transcript confidence alone is insufficient to justify an action, so better listening alone will not remove these failures (Section~\ref{sec:limits}).
\item \textbf{A reference architecture} that separates consequential action control from perception and reasoning. Its decision layer chooses to act, wait, ask, or remain silent based on addressedness, speaker identity, content confidence, tool reliability, utterance completeness, and action risk. Tools report their reliability through an evidence contract, and fuzzy inference, evolutionary optimization, and sequential decision-making implement the layer (Section~\ref{sec:arch}).
\item \textbf{Empirical support} for each element of the architecture from controlled simulations (200{,}000 episodes, each a single proposed action), leave-one-input-out ablation studies, and recorded audio from two public benchmarks, including egocentric recordings (Section~\ref{sec:evidence}). On recorded audio, explicit decision gating reduces executed side talk from 41\% to 7--10\% while serving most valid requests, outperforming prompting-based approaches.
\item \textbf{Implications for designers, decision-level evaluation criteria, and a research agenda} for building and evaluating such architectures (Section~\ref{sec:action}).
\end{enumerate}
\section{The Decision Gap in Audio Agents}\label{sec:limits}

\subsection{How Far Listening Has Come}\label{sec:great}

Audio intelligence has moved from task-specific statistical models to end-to-end deep networks. Weakly supervised training at scale produced recognizers such as Whisper that are robust to noise, accents, and domain shift \cite{radford2023whisper}. Foundation models then cast audio generation as language modeling over discrete tokens \cite{borsos2022audiolm, wang2023valle}, LALMs coupled audio encoders with language models to understand speech, sounds, and music in one system \cite{tang2023salmonn, chu2024qwen2, rubenstein2023audiopalm}, and spoken dialogue models progressed from turn-based exchanges \cite{zhang2023speechgpt, fang2024llama} to real-time, full-duplex conversation \cite{zhang2025omniflatten, kyutai2024moshi}. In parallel, agent research showed that language models can interleave reasoning with tool calls \cite{yao2023react}, learn when to invoke tools \cite{schick2024toolformer}, and accumulate skills over time \cite{wang2023voyager}. Audio agents adapt these ideas to sound and follow three design paradigms, compared in Table~\ref{tab:systems}.

\textit{Text-orchestrated agents} use a text LLM that plans and calls audio models as tools without hearing the signal itself. AudioGPT \cite{huang2023audiogpt} connects an LLM to audio foundation models, Speech-Copilot \cite{kuan2024speechcopilot} decomposes instructions into speech toolsets through program generation, and AudioAgent \cite{wang2024audio} plans audio generation and editing. AudioToolAgent \cite{wijngaard2025audiotoolagent} coordinates several LALMs and a speech recognizer, cross-checks their outputs, and reaches 77.50\% accuracy on MMAU and 77.00\% on MMAR, two audio understanding and reasoning benchmarks. Audio-Mind \cite{wang2026audiomind} places an LALM in front of a text planner and organizes tools by reliability, arguing that as LALMs improve, the key question shifts from enabling tool use to deciding when additional evidence is worth acquiring.

\textit{Audio-native tool users} let the reasoning model hear the audio and call tools when needed. Audio-Maestro \cite{lee2025audio} adds timestamped tools such as chord recognition, diarization, and emotion detection, and SpeechAgent-R \cite{wang2026hiu} learns, through multi-turn reinforcement learning on 65,492 trajectories across 24 tasks, when to invoke its 8 skills and 9 tools. \textit{End-to-end conversational agents} such as OmniFlatten \cite{zhang2025omniflatten} listen and speak simultaneously, supporting interruptions and backchannels, but generally do not call external tools. Beyond these paradigms, embodied systems connect audio to physical action: ROSA \cite{royce2025enabling} translates natural-language commands, including spoken ones, into Robot Operating System calls with parameter validation and constraint enforcement, one of the few designs with explicit action gating.

\begin{table*}[t]
\centering
\caption{Representative audio agents compared on how they listen and which of the six decisions of Table~\ref{tab:failures} they explicitly support. ``Hears audio'' means that the component that chooses actions receives audio representations rather than only text. \cmark: explicit mechanism described by the authors; \pmark: partial, for example used to verify answers rather than to gate actions, or timing for speech but not for actions; \xmark: none described.}
\label{tab:systems}
\renewcommand{\arraystretch}{1.15}
\small
\setlength{\tabcolsep}{4pt}
\begin{tabularx}{\textwidth}{P{2.9cm} L C{1.0cm} C{0.9cm} C{1.35cm} C{1.05cm} C{0.9cm} C{1.35cm} C{0.9cm} C{0.9cm}}
\toprule
& & & & \multicolumn{6}{c}{\textbf{Decisions supported}} \\
\cmidrule(lr){5-10}
\textbf{System} & \textbf{Paradigm} & \textbf{Hears audio} & \textbf{Uses tools} & \textbf{\footnotesize Whether} & \textbf{\footnotesize Whom} & \textbf{\footnotesize What} & \textbf{\footnotesize Which\newline evidence} & \textbf{\footnotesize When} & \textbf{\footnotesize Safe} \\
\midrule
AudioGPT \cite{huang2023audiogpt} & Text-orchestrated & \xmark & \cmark & \xmark & \xmark & \xmark & \xmark & \xmark & \xmark \\
Speech-Copilot \cite{kuan2024speechcopilot} & Text-orchestrated & \xmark & \cmark & \xmark & \xmark & \xmark & \xmark & \xmark & \xmark \\
AudioAgent \cite{wang2024audio} & Text-orchestrated & \xmark & \cmark & \xmark & \xmark & \xmark & \xmark & \xmark & \xmark \\
AudioToolAgent \cite{wijngaard2025audiotoolagent} & Text-orchestrated & \xmark & \cmark & \xmark & \xmark & \pmark & \pmark & \xmark & \xmark \\
Audio-Mind \cite{wang2026audiomind} & Text planner with LALM front end & \pmark & \cmark & \xmark & \xmark & \xmark & \cmark & \xmark & \xmark \\
Audio-Maestro \cite{lee2025audio} & Audio-native tool user & \cmark & \cmark & \xmark & \xmark & \xmark & \xmark & \xmark & \xmark \\
SpeechAgent-R \cite{wang2026hiu} & Audio-native tool user & \cmark & \cmark & \xmark & \xmark & \xmark & \pmark & \xmark & \xmark \\
OmniFlatten \cite{zhang2025omniflatten} & End-to-end full-duplex & \cmark & \xmark & \xmark & \xmark & \xmark & \xmark & \pmark & \xmark \\
ROSA \cite{royce2025enabling} & Embodied middleware & \pmark & \cmark & \xmark & \xmark & \xmark & \xmark & \xmark & \pmark \\
\midrule
\textbf{Proposed (Section~\ref{sec:arch})} & Decision layer behind any of the above & \cmark & \cmark & \cmark & \cmark & \cmark & \cmark & \cmark & \cmark \\
\bottomrule
\end{tabularx}
\end{table*}

Beyond individual systems, Table~\ref{tab:taxonomy} groups agentic audio systems into conversational, tool-augmented, collaborative, embodied, and ambient settings. Deployments range from mental-health conversational agents \cite{lee2025artificial} and meeting assistants \cite{senthilselvi2025real} to health monitoring from speech and cough \cite{wang2024sound, deshpande2022ai}, smart homes \cite{shabbir2025enhancing}, proactive wearable assistants \cite{mahfuz2026proactive}, and public safety \cite{omarov2025dangerous, dominguez2025real}, increasingly fused with non-acoustic sensors \cite{srivastava2026beyond, doukas2010emergency}. Wearable agents, in particular, must distinguish ambient sound, conversations not directed at them, and speech directed at them \cite{fung2025embodied}. As agents move from answering questions to controlling devices and calling emergency services, the cost of a wrong action rises sharply, as the last two columns of Table~\ref{tab:taxonomy} show.

\begin{table*}[t]
\centering
\caption{Our taxonomy of agentic audio systems, with the actions each class takes and its dominant decision risk. Embodiment ranges from disembodied software, through soft-bodied agents with a virtual or limited physical presence such as wearables, to physically embodied robots; autonomy ranges from reactive (fixed responses) and interactive (turn-by-turn) to deliberative (multi-step planning) and autonomous (acting without per-step confirmation).}
\label{tab:taxonomy}
\renewcommand{\arraystretch}{1.12}
\small
\begin{tabularx}{\textwidth}{P{2.3cm} P{2.0cm} P{1.9cm} P{2.0cm} L L}
\toprule
\textbf{Category} & \textbf{Interaction modality} & \textbf{Embodiment} & \textbf{Autonomy level} & \textbf{Typical consequential actions} & \textbf{Dominant decision risk} \\
\midrule
Conversational audio agents & Speech-centric to full audio & Disembodied to soft-bodied & Interactive to deliberative & Bookings, messages, purchases, account changes & Acting on misheard content; acting before a correction (what, when) \\
Tool-augmented audio systems & Full audio & Disembodied & Deliberative to autonomous & Analyses, reports, downstream automation & Trusting unreliable or conflicting tool outputs (which evidence) \\
Collaborative multi-agent systems & Full audio to multimodal & Disembodied to soft-bodied & Deliberative to autonomous & Shared plans, delegated sub-tasks & Errors propagating between agents without a shared confidence signal \\
Embodied audio agents & Speech-centric to multimodal & Physically embodied & Deliberative to autonomous & Motion, manipulation, device control & Following a bystander; acting on ambiguous references (whether, whom) \\
Ambient audio agents & Environmental audio & Disembodied to embedded & Reactive to deliberative & Alerts, emergency calls, logging & False alarms and missed events; recording bystanders (whether, safe) \\
\bottomrule
\end{tabularx}
\end{table*}

Perception benchmarks show fast progress \cite{sakshi2025mmau, wijngaard2025audiotoolagent, wang2026audiomind}, with two caveats that matter for decisions. First, standardized evaluation reveals significant gaps in temporal understanding, including diarization, so ``who spoke when'' remains difficult \cite{surapaneni2025harness}. Second, LALMs rely on lexical content more than on acoustic cues, predicting neutral emotion from neutral words even when the voice is not neutral \cite{chen2026listen}. Audio agents therefore recover \emph{what} was said far better than \emph{who} said it, \emph{how}, and \emph{to whom}, which are exactly the cues on which decisions depend.

\subsection{Prior Work on Knowing When Not to Act}\label{sec:sds}\label{sec:textagents}

Deciding whether to act on uncertain speech is not a new problem. Spoken dialogue systems have long used recognition confidence to accept, confirm, or reject a request, and have studied how to recover when understanding fails \cite{bohus2005sorry}. Statistical dialogue managers went further, maintaining a belief over the user's intent and selecting actions, including confirmations, by expected reward in a partially observable Markov decision process \cite{williams2007pomdp, young2013pomdp}. Device-directed speech detection asks whether an utterance was meant for the system at all \cite{shriberg2012learning, mallidi2018devicedirected}, and turn-taking models decide when the user has finished \cite{raux2009finite, skantze2021turn}. These systems, however, served a single user in a turn-based exchange, relied on a single recognizer, and mapped speech to slot values that the next turn could correct.

For text agents, the question of when not to act is now studied intensively. AgentAbstain pairs each task with a minimally perturbed variant that should not be executed, so that no constant policy can exceed 50\%. The best of 17 frontier models reaches only 59.5\% paired accuracy (both variants handled correctly), act accuracy exceeds abstain accuracy by 21 points on average, and some agents commit irreversible actions before recognizing that they should have stopped \cite{liu2026agentabstain}. Agents also abstain too late \cite{luo2026agenticabstention} and rarely ask for help when instructions are incomplete \cite{trinh2026hilbench}, while benchmarks that reward task completion ignore abstention by construction \cite{ojewale2026abstention}. Conformal methods give language-model planners statistical guarantees on when to ask \cite{ren2023knowno}.

Agentic audio AI departs from both settings in four ways. First, \emph{addressedness becomes a decision variable}: every typed message is addressed to a text agent, whereas most sound that reaches a microphone is not. Second, \emph{uncertainty originates in the signal and its source}, not only in the instruction: a well-specified command can be misheard, spoken by a bystander, or played by a television. Third, \emph{evidence arrives incrementally}, so the agent must decide when to commit while the utterance is still unfolding and corrections may follow. Fourth, \emph{actions are tool calls} with graded and often irreversible effects, such as payments, messages, and device control, informed by several tools of varying reliability. Neither classical dialogue management nor text-agent abstention therefore transfers directly.

Our contribution thus lies not in confirmation, abstention, or belief-based decision-making individually, but in integrating them at the boundary between reasoning and consequential action. As Section~\ref{sec:theory} argues, current audio agents have largely abandoned the belief and confirmation machinery of classical dialogue managers. The decision layer of Section~\ref{sec:arch} restores this machinery and extends it with evidence on addressedness, speaker identity, tool reliability, utterance completeness, and action risk, so that uncertainty-aware dialogue management carries over to tool-augmented agents operating in open acoustic environments.

\subsection{Six Decision Failures}\label{sec:poor}

\begin{figure}[t]
\centering
\includegraphics[width=\columnwidth]{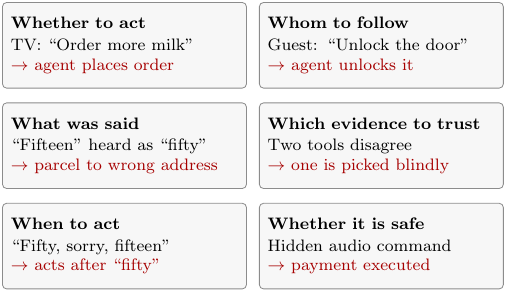}
\caption{Six decision failures of audio agents, each shown as an audio scenario and the wrong action that follows. Table~\ref{tab:failures} gives the supporting evidence for each.}
\label{fig:failures}
\end{figure}

An agent that turns audio into action must answer six questions, explicitly or not: whether to act, whom to follow, what was said, which evidence to trust, when to act, and whether the action is safe. Fig.~\ref{fig:failures} illustrates each failure, and Table~\ref{tab:failures} lists its evidence, remedy, and candidate CI techniques. The failures persist even though relevant components exist. Speaker verification \cite{desplanques2020ecapa} and diarization \cite{lee2025audio} inform answers rather than gate actions. Separation models \cite{luo2019convtasnet} and event taggers trained on AudioSet \cite{gemmeke2017audioset, kong2020panns} can mislead downstream tools without reporting their reliability. Full-duplex models handle timing for speech \cite{kyutai2024moshi} but not for actions, and self-corrections remain among their most consistent tool-calling failures \cite{lin2026fdbv3}. Voice assistants can also be hijacked by inaudible commands \cite{yan2019feasibility}, and rule-based safeguards \cite{royce2025enabling} are rarely combined with acoustic evidence about who is speaking.

\begin{table*}[t]
\centering
\caption{Six decision failures of audio agents, with the evidence for each, the remedy proposed in Section~\ref{sec:arch}, and the CI techniques that can implement it.}
\label{tab:failures}
\renewcommand{\arraystretch}{1.12}
\footnotesize
\begin{tabularx}{\textwidth}{P{1.7cm} P{3.1cm} L P{3.4cm} P{3.2cm}}
\toprule
\textbf{Decision} & \textbf{Failure mode} & \textbf{Evidence} & \textbf{Remedy} & \textbf{CI technique} \\
\midrule
Whether to act & Executes commands meant for someone else, self-talk, or media & Tool chosen in 96--97\% of controls, silence in 1\% of speaker switches \cite{zhang2026vgbench}; directed vs.\ non-directed speech \cite{fung2025embodied} & Explicit \textsc{silent} action; addressedness input & Fuzzy action gate; reinforcement learning with a silence action \\
Whom to follow & Attributes a bystander's words to the user & Bystander substitution under competing speech \cite{ivry2026voice} & Speaker match as a gating input & Fuzzy rules on speaker-match degree \\
What was said & Acts on misheard or never-spoken words; misses tone & Substitutions and hallucinated words under noise and clipping \cite{ivry2026voice}; lexical dominance \cite{chen2026listen} & Audio kept in the loop; word-level confidence in the evidence contract & Calibration \cite{guo2017calibration}; selective prediction \cite{geifman2017selective} \\
Which evidence to trust & Resolves tool conflicts arbitrarily & Agent arbitrates tools without hearing the audio \cite{wijngaard2025audiotoolagent}; tools reshape the signal \cite{wang2026audiomind} & Reliability-aware tools and evidence fusion & Reliability-weighted fuzzy fusion; neuro-fuzzy learning \cite{jang1993anfis} \\
When to act & Commits before corrections, or waits too long & ``Fifty, sorry, fifteen'' example \cite{ivry2026voice}; turn-taking handled only for speech \cite{zhang2025omniflatten, ekstedt2022vap} & Late commitment for irreversible actions & Reinforcement learning with latency and wrong-action penalties \cite{sutton2018rl} \\
Whether it is safe & Follows hidden or malicious audio instructions; records bystanders & Narrative audio attacks \cite{yu2026narrative}; defenses lag attacks \cite{luo2026lalmsurvey} & Risk-aware gating, confirmation, on-device filtering & Fuzzy risk assessment; asking under uncertainty \cite{ren2023knowno} \\
\bottomrule
\end{tabularx}
\end{table*}

\subsection{Why Better Listening Will Not Fix It}\label{sec:theory}

\textit{A common root cause.} The six failures share one cause. In current designs, the path from hearing to acting runs through a reasoning model that outputs an action directly. Unlike a classical dialogue manager, such a model keeps no explicit belief over what was meant: uncertainty from perception and tools is not carried forward, the cost of a wrong action is not represented, and \textsc{wait}, \textsc{ask}, and \textsc{silent} are not first-class choices. Better perception alone cannot fix this, because even a perfect recognizer cannot tell whether a correctly recognized command was meant for the agent. Nor can scale alone: in text agents, abstention ability is largely independent of task-solving ability \cite{liu2026agentabstain}. The remedy is architectural, and a simple decision model makes its logic precise.

\textit{A decision model.} Consider an agent that has heard an audio stream $x_{1:t}$ and whose reasoning core proposes an action $a$. The action is correct only if three conditions hold jointly: the speech was addressed to the agent ($z=1$), it came from a speaker authorized to request $a$ ($u=1$), and its content was recovered correctly ($c=1$). Let
\begin{equation}
p = P(z=1,\, u=1,\, c=1 \mid x_{1:t})
\label{eq:pok}
\end{equation}
be the probability that executing $a$ is what an authorized user intended. The agent chooses a decision $d \in \{\textsc{act}, \textsc{wait}, \textsc{ask}, \textsc{silent}\}$. A correct action yields gain $G$, and a wrong one costs $C(a)$, which grows with irreversibility. Asking costs the user effort $\kappa$, and remaining silent on a genuine request costs $M$.

\textit{A cost-aware action threshold.} The following result adapts Chow's optimal reject rule \cite{chow1970reject}, in which a classifier abstains when its posterior falls below a cost-dependent threshold, to an agent whose alternative to acting is a question. It compares acting with asking only, and therefore gives a cost-sensitive threshold on $p$ rather than a complete policy over all four decisions.

\textit{Proposition~1.} Consider a request addressed to the agent, and assume that asking resolves the remaining uncertainty and yields gain $G$ at effort cost $\kappa$. Acting is preferred to asking if and only if
\begin{equation}
p \;\geq\; \tau(a) = 1 - \frac{\kappa}{G + C(a)}.
\label{eq:tau}
\end{equation}

\textit{Proof.} The expected utility of acting is $pG - (1-p)C(a)$ and that of asking is $G-\kappa$. The first is at least the second exactly when $(1-p)\,(G + C(a)) \leq \kappa$. \hfill$\square$

\begin{figure}[t]
\centering
\includegraphics[width=\columnwidth]{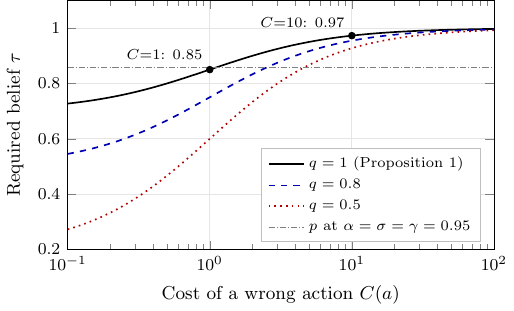}
\caption{Action threshold $\tau_q(a)$ versus the cost of a wrong action, for $G=1$, $\kappa=0.3$, $M=0$, and clarification success probability $q$. The threshold approaches 1 as actions become costly. The dash-dotted line marks the belief $p\approx0.86$ attainable when addressedness, speaker, and content confidence are each $0.95$; it clears the threshold only for actions with $C(a) \lesssim 1.1$ (evidence bound).}
\label{fig:threshold}
\end{figure}

Questions do not always succeed: users mishear, answer vaguely, or leave. If a question resolves the request with probability $q$ and the agent otherwise remains silent, asking yields $qG-(1-q)M-\kappa$, and the threshold becomes $\tau_q(a) = 1 - \big((1-q)(G+M) + \kappa\big)/\big(G + C(a)\big)$, which reduces to \eqref{eq:tau} when $q=1$. Less reliable clarification lowers the threshold, but for costly actions $\tau_q$ stays close to 1. In both cases the threshold rises with the cost of being wrong (Fig.~\ref{fig:threshold}). With $G=1$ and $\kappa=0.3$, a reversible action with $C=1$ requires $p \geq 0.85$, whereas an irreversible one with $C=10$ requires $p \geq 0.97$. The same evidence can therefore justify reading the weather aloud but not placing an order, which is the formal basis for risk-aware gating (Section~\ref{sec:principles}).

\textit{An evidence bound.} The threshold fixes how much belief an action needs; the evidence caps how much is available. Because the action is correct only if $z=1$, $u=1$, and $c=1$ all hold, $p \leq \min(\alpha, \sigma, \gamma)$, with $\alpha = P(z=1 \mid x_{1:t})$, $\sigma = P(u=1 \mid x_{1:t})$, and $\gamma = P(c=1 \mid x_{1:t})$. This is the elementary Fr\'echet upper bound on a joint probability, and it holds without any independence assumption. If the three conditions are, in addition, conditionally independent given the audio, then
\begin{equation}
p = \underbrace{P(z=1 \mid x_{1:t})}_{\alpha}\;\underbrace{P(u=1 \mid x_{1:t})}_{\sigma}\;\underbrace{P(c=1 \mid x_{1:t})}_{\gamma},
\label{eq:decomp}
\end{equation}
which is how we estimate $p$ in practice; Section~\ref{sec:sim} tests this estimator under dependence. Here, speaker verification serves as a proxy for authorization under an enrollment-based access policy in which the enrolled user may request the actions considered; $\sigma$ is therefore not an assessment of action-specific permissions.

Under either form, transcript confidence $\gamma$ alone is insufficient to justify an action. Even $\alpha=\sigma=\gamma=0.95$ gives $p \approx 0.86$, which, with the costs above and $q=1$, falls below the threshold of every action with $C > 1.1$ (Fig.~\ref{fig:threshold}). Text agents implicitly assume $\alpha=\sigma=1$, which is why their abstention methods do not transfer to audio without modification; we refer to this result as the \emph{evidence bound}. Acting on it requires quantities that current systems do not represent: calibrated estimates of $\alpha$, $\sigma$, and $\gamma$, and action costs $C(a)$. In practice, these quantities are also imperfect: model confidences are often miscalibrated \cite{guo2017calibration}, and costs are typically known only in linguistic terms such as ``hard to undo.'' This motivates the architecture of the next section, whose decision layer holds exactly these quantities and is built with methods that tolerate graded, imprecise inputs.

\section{Proposed Architecture}\label{sec:arch}

This section closes the decision gap with a reference architecture and its design principles, an evidence contract through which tools report their reliability, and CI methods that implement the decision layer.

\subsection{Architecture Overview}\label{sec:archdesc}

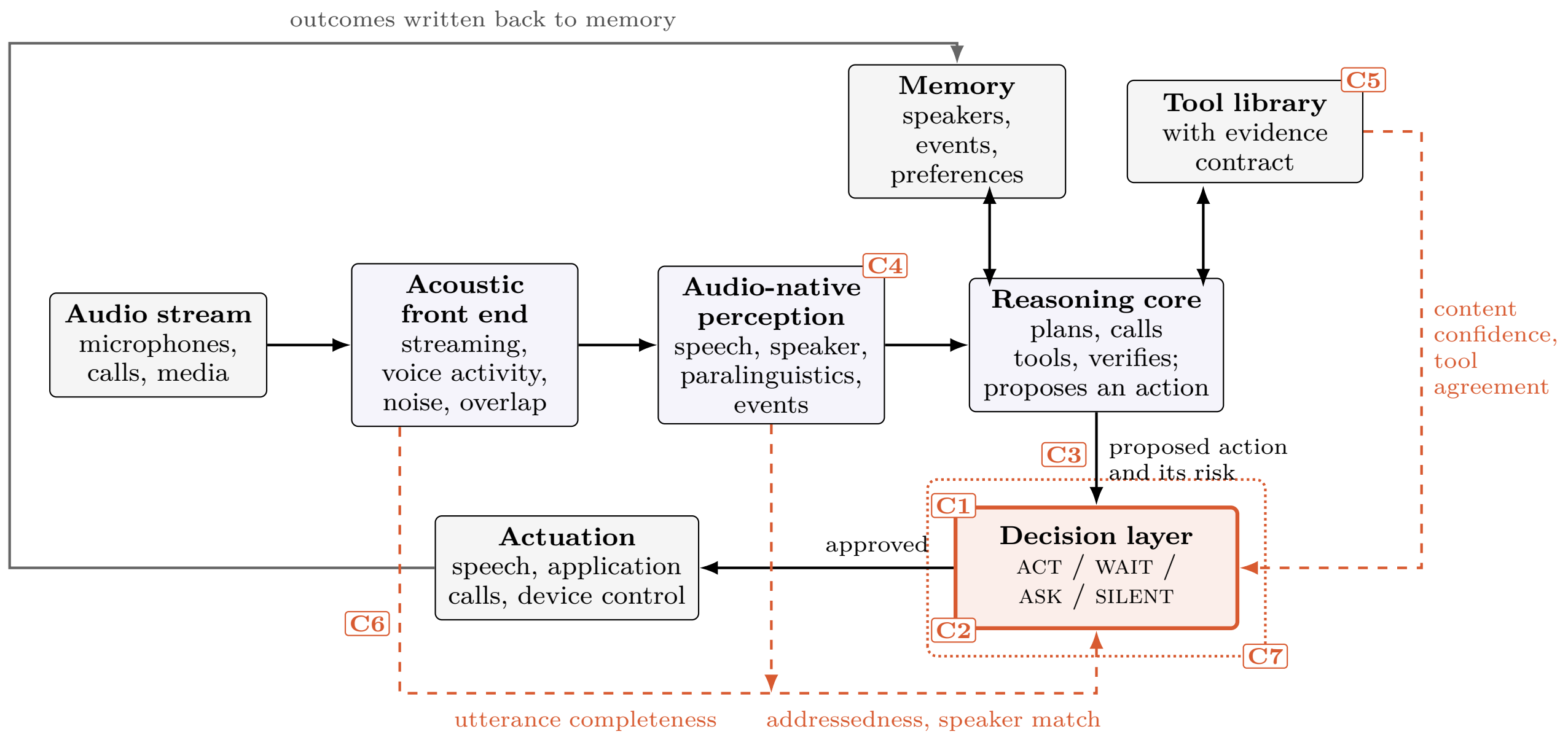
\begin{figure*}[t]
\centering
\begin{tikzpicture}[
  font=\footnotesize,
  blk/.style={draw, rounded corners=2pt, align=center, minimum height=12mm, text width=22mm, fill=corecolor!6},
  side/.style={draw, rounded corners=2pt, align=center, minimum height=11mm, text width=21mm, fill=black!4},
  gate/.style={draw=gatecolor, very thick, rounded corners=2pt, align=center, minimum height=13mm, text width=28mm, fill=gatecolor!10},
  arr/.style={-{Latex[length=2mm]}, thick},
  darr/.style={{Latex[length=2mm]}-{Latex[length=2mm]}, thick},
  evid/.style={-{Latex[length=2mm]}, thick, dashed, draw=gatecolor},
  tag/.style={draw=gatecolor, fill=white, text=gatecolor, rounded corners=1pt, inner sep=1.2pt, font=\scriptsize\bfseries}
]
\node[side] (in) at (0,0) {\textbf{Audio stream}\\ microphones, calls, media};
\node[blk] (fe) at (3.3,0) {\textbf{Acoustic front end}\\ streaming, voice activity, noise, overlap};
\node[blk] (per) at (6.6,0) {\textbf{Audio-native perception}\\ speech, speaker, paralinguistics, events};
\node[blk, text width=25mm] (core) at (10.1,0) {\textbf{Reasoning core}\\ plans, calls tools, verifies; proposes an action};
\node[side] (mem) at (8.6,2.3) {\textbf{Memory}\\ speakers, events, preferences};
\node[side, text width=23mm] (tools) at (11.7,2.3) {\textbf{Tool library}\\ with evidence contract};
\node[gate] (gate) at (10.1,-2.4) {\textbf{Decision layer}\\ \textsc{act} / \textsc{wait} / \textsc{ask} / \textsc{silent}};
\node[side, text width=26mm] (act) at (4.4,-2.4) {\textbf{Actuation}\\ speech, application calls, device control};

\draw[gatecolor, densely dotted, thick, rounded corners=3pt]
  ($(gate.south west)+(-0.28,-0.28)$) rectangle ($(gate.north east)+(0.28,0.28)$);

\draw[arr] (in) -- (fe);
\draw[arr] (fe) -- (per);
\draw[arr] (per) -- (core);
\draw[darr] (8.95,1.72) -- (8.95,0.62);
\draw[darr] (11.25,1.72) -- (11.25,0.62);
\draw[arr] (core) -- node[right, align=left, font=\scriptsize]{proposed action\\ and its risk} (gate);
\draw[arr] (gate) -- node[pos=0.3, above, font=\scriptsize]{approved} (act);
\draw[arr, draw=black!60] (act.west) -- (-1.6,-2.4) -- (-1.6,3.25)
  -- node[pos=0.5, above, font=\scriptsize, text=black!70]{outcomes written back to memory} (8.6,3.25) -- (mem.north);
\draw[evid] ($(fe.south)+(-0.7,0)$) -- (2.6,-3.75) -- (10.1,-3.75) -- (gate.south);
\draw[evid] (per.south) -- (6.6,-3.75);
\node[font=\scriptsize, text=gatecolor, align=center] at (4.6,-4.05) {utterance completeness};
\node[font=\scriptsize, text=gatecolor, align=center] at (8.35,-4.05) {addressedness, speaker match};
\draw[evid] (tools.east) -- (13.6,2.3) -- node[pos=0.5, right, font=\scriptsize, text=gatecolor, align=left]{content\\ confidence,\\ tool\\ agreement} (13.6,-2.4) -- (gate.east);
\node[tag] at (gate.north west) {C1};
\node[tag] at (gate.south west) {C2};
\node[tag] at (9.75,-1.18) {C3};
\node[tag] at (per.north east) {C4};
\node[tag] at (tools.north east) {C5};
\node[tag] at (2.25,-3.0) {C6};
\node[tag] at ($(gate.south east)+(0.28,-0.28)$) {C7};
\end{tikzpicture}
\caption{Proposed reference architecture for agentic audio AI. Gray and purple blocks are components common to current audio agents; orange elements are our additions: the decision layer, the evidence paths that feed it (dashed), and the evidence contract on the tool library. Actuation writes the outcomes of executed actions back to memory. Tags C1--C7 mark the claim tested about each element, placed on the component or evidence path it concerns; C7 concerns the decision layer as a whole and is drawn as the dotted block around it.}
\label{fig:arch}
\end{figure*}

Fig.~\ref{fig:arch} shows the proposed reference architecture. Its central addition is a \emph{decision layer} between the reasoning core and actuation: the reasoning core no longer executes actions but proposes them, and the decision layer chooses to act, wait, ask, or remain silent (Section~\ref{sec:ci}). The layer receives the proposed action and its risk, together with evidence on addressedness, speaker match, content confidence, tool agreement, and utterance completeness, delivered along the dashed paths and by tools that report their reliability through an evidence contract (Section~\ref{sec:contract}). These additions, shown in orange, are our contributions; the remaining components are common to current audio agents.

The \emph{acoustic front end} performs streaming capture, voice activity detection, and optional multichannel processing, and reports scene cues such as signal-to-noise ratio and overlap. \emph{Audio-native perception} builds representations of speech, speakers, paralinguistic cues, and events directly from the signal \cite{tang2023salmonn, chu2024qwen2}. The \emph{reasoning core} interprets the goal, plans, calls tools, and verifies results in the spirit of ReAct \cite{yao2023react}, but outputs only a \emph{proposed} action. The \emph{tool library} registers tools under the evidence contract; \emph{memory} holds the dialogue, the audio buffer, speaker profiles, and timestamped events; and \emph{actuation} executes approved actions and writes their outcomes back to memory.

Perception must also supply two estimates that are open problems in themselves. \emph{Addressedness} $\alpha$ is the task of device-directed speech detection: early detectors using acoustic, lexical, or recognition features in isolation reached equal error rates of 9--20\% \cite{mallidi2018devicedirected}, and multichannel spatial cues raise side-talk rejection substantially \cite{lin2026wearvox}. \emph{Speaker match} $\sigma$ comes from speaker verification \cite{desplanques2020ecapa} and diarization \cite{lee2025audio}, both of which degrade with short utterances and overlap. Neither estimate will be perfect, so the decision layer must reason about their uncertainty.

\subsection{Design Principles}\label{sec:principles}

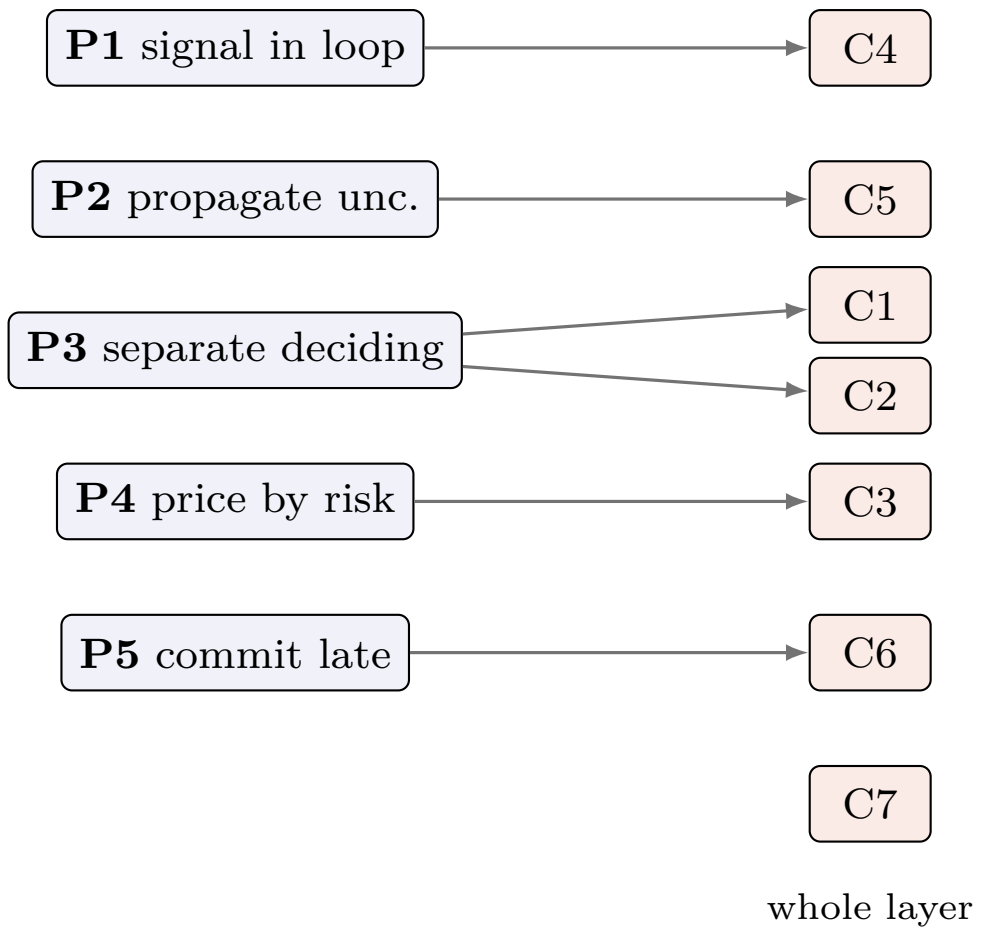
\begin{figure}[t]
\centering
\begin{tikzpicture}[
  font=\footnotesize,
  pnode/.style={draw, rounded corners=2pt, minimum width=22mm, minimum height=5mm, align=center, fill=corecolor!8},
  cnode/.style={draw, rounded corners=2pt, minimum width=8mm, minimum height=5mm, fill=gatecolor!12},
  link/.style={-{Latex[length=1.5mm]}, semithick, draw=black!55}
]
\node[pnode] (P1) at (0,4) {\textbf{P1} signal in loop};
\node[pnode] (P2) at (0,3) {\textbf{P2} propagate unc.};
\node[pnode] (P3) at (0,2) {\textbf{P3} separate deciding};
\node[pnode] (P4) at (0,1) {\textbf{P4} price by risk};
\node[pnode] (P5) at (0,0) {\textbf{P5} commit late};
\foreach \i/\y in {4/4.0,5/3.0,1/2.3,2/1.7,3/1.0,6/0.0}
  \node[cnode] (C\i) at (4.2,\y) {C\i};
\node[cnode] (C7) at (4.2,-1.0) {C7};
\draw[link] (P1) -- (C4);
\draw[link] (P2) -- (C5);
\draw[link] (P3) -- (C1);
\draw[link] (P3) -- (C2);
\draw[link] (P4) -- (C3);
\draw[link] (P5) -- (C6);
\node[align=left, font=\scriptsize] at (4.2,-1.7) {whole layer};
\end{tikzpicture}
\caption{Each design principle (P1--P5) is tested by one or two claims; claim C7 concerns the decision layer as a whole rather than a single principle. Section~\ref{sec:evidence} reports the evidence for each claim.}
\label{fig:pc}
\end{figure}

The orange elements of Fig.~\ref{fig:arch} follow from five design principles, each paired with a claim tested in Section~\ref{sec:evidence}; Fig.~\ref{fig:pc} summarizes the links.

\begin{enumerate}
\item[\textbf{P1}] \textbf{Keep the signal in the loop.} The reasoning core should be able to re-listen to the audio. Transcripts, captions, and labels are tool outputs, not the only interface to sound. \textbf{Claim C4:} addressedness judged from the audio is more reliable than addressedness judged from the transcript. (C4 concerns the \emph{source} of the addressedness estimate, whereas C2 concerns its \emph{effect} on wrong actions.)
\item[\textbf{P2}] \textbf{Propagate uncertainty.} Every tool returns not only an output but also a confidence and the conditions under which that output is valid. \textbf{Claim C5:} weighting tools by confidence and validity yields better content evidence than treating them equally.
\item[\textbf{P3}] \textbf{Separate deciding from reasoning.} The choice to act is made by a dedicated, auditable decision layer that receives evidence about who spoke and to whom, rather than implicitly inside the reasoning model. \textbf{Claim C1:} a separate decision layer outperforms letting the model act directly. \textbf{Claim C2:} addressedness and speaker evidence prevent more wrong actions than transcript confidence.
\item[\textbf{P4}] \textbf{Price actions by risk.} Reversible, low-cost actions need less evidence than irreversible or safety-critical ones. \textbf{Claim C3:} pricing actions by risk keeps costly errors rare, even under miscalibration.
\item[\textbf{P5}] \textbf{Stream, and commit late.} Decisions are revisable until an action is committed, and commitment for risky actions waits for utterance completion. \textbf{Claim C6:} waiting before risky actions avoids acting on self-corrected words at a small latency cost.
\end{enumerate}

A seventh claim, C7, concerns the decision layer as a whole: because the architecture prescribes what the layer receives but not how it is implemented, well-tuned implementations should perform alike.

\subsection{Tools Reporting Reliability}\label{sec:contract}

Building on the reliability-aware view of tools in Audio-Mind \cite{wang2026audiomind}, Table~\ref{tab:tools} organizes callable tools by what they return and when they fail; its last column, our addition, gives the natural reliability signal of each category. We make this signal explicit through an \emph{evidence contract}. A call to tool $k$ on audio segment $x$ returns a triple
\begin{equation}
\mathcal{T}_k(x) = \big(y_k,\; c_k,\; v_k\big),
\label{eq:contract}
\end{equation}
where $y_k$ is the output, $c_k \in [0,1]$ is a calibrated confidence \cite{guo2017calibration}, and $v_k \in [0,1]$ is a validity degree that measures how well the current acoustic conditions match those in which the tool is reliable, such as its usable signal-to-noise ratio, overlap, language, or duration. Audio LLMs can be trained to describe the quality of the speech they receive \cite{chen2025audio}, a direct source of validity estimates. The validity degree makes explicit what is usually implicit: a recognizer is less trustworthy in overlapping speech, and an event tagger outside its training ontology. Each triple travels with its audio segment $s_k$ and the condition features $\mathbf{e}_k$ from which $v_k$ was computed, so that every decision can be traced back to its evidence.

On a single GPU, the fuzzy gate takes under 0.1~ms, so latency is dominated by computing the evidence. Speaker verification (12~ms) and the independent addressedness detector of Section~\ref{sec:real} (9~ms) fit within a 100~ms budget after the last tool output, roughly the limit below which a spoken response feels immediate; querying the agent's own LALM about addressedness (142~ms median, 231~ms at the 95th percentile) does not. Addressedness is therefore best estimated outside the reasoning core.

\textit{Integration with full-duplex models.} The decision layer does not slow conversation. A full-duplex model may keep generating backchannels and responses token by token, but any tool call it emits is held at the decision layer until approved. Commitment thus happens at the tool-call boundary, where irreversibility arises, rather than at the token level. The same holds for cascades and audio-native tool users: the layer needs only the proposed tool call, its risk, and the evidence of \eqref{eq:contract}.

\begin{table*}[t]
\centering
\caption{Reliability-aware taxonomy of tools that audio agents can call, with what each returns.}
\label{tab:tools}
\renewcommand{\arraystretch}{1.1}
\footnotesize
\begin{tabularx}{\textwidth}{P{5.6cm} L P{4.4cm}}
\toprule
\textbf{Category: examples (returns)} & \textbf{Typical failure conditions} & \textbf{Reliability signal for the decision layer} \\
\midrule
Signal operators: enhancement, separation \cite{luo2019convtasnet} (modified audio) & Artifacts that mislead downstream tools; over-suppression of quiet speech & Agreement of results on raw and processed audio \\
Segmentation and localization: voice activity, sound events \cite{kong2020panns}, diarization \cite{lee2025audio} (segments, labels) & Overlap, short segments, events outside the training ontology \cite{gemmeke2017audioset} & Segment-level posteriors; overlap ratio \\
Speech content: recognition \cite{radford2023whisper} (words with timestamps) & Noise, clipping, accents, rare names and numbers \cite{ivry2026voice} & Word-level confidence; agreement across recognizers \\
Speaker and paralinguistics: verification \cite{desplanques2020ecapa}, emotion \cite{lee2025audio} (identity scores, affect labels) & Short utterances, channel mismatch, conflicting lexical and acoustic cues \cite{chen2026listen} & Speaker-match score; lexical--acoustic agreement \\
Scene and audio reasoning: LALMs as tools \cite{tang2023salmonn, chu2024qwen2, wijngaard2025audiotoolagent} (free text, captions) & Hallucination, lexical bias, long recordings & Self-consistency across queries; cross-model agreement \\
Generation: text-to-speech \cite{wang2023valle}, text-to-audio \cite{liu2023audioldm} (new audio) & Mispronunciation; misuse for impersonation & Automatic quality checks \cite{chen2025audio}; consent and provenance flags \\
External actions: application interfaces, search, robot control \cite{royce2025enabling} (effects in the world) & Irreversible side effects; authorization errors & Action risk and reversibility score \\
\bottomrule
\end{tabularx}
\end{table*}

\subsection{Building the Decision Layer with Computational Intelligence}\label{sec:ci}

The decision layer must combine graded evidence, respect the cost of errors, remain auditable, and adapt with limited data, all strengths of CI. The last column of Table~\ref{tab:failures} maps each technique to the decisions it supports. Section~\ref{sec:evidence} evaluates the four components below, and Section~\ref{sec:challenges} discusses learning the gate from logged outcomes with statistical guarantees.

\textit{Silence and waiting.} Asking is worthwhile only if someone is likely to be addressing the agent. With addressedness probability $\alpha$, asking has expected utility $\alpha G - \kappa$ and silence $-\alpha M$, so asking is preferred only when $\alpha > \kappa/(G+M)$; in practice, designers set a higher bar, because questioning a bystander has social costs beyond $\kappa$. Waiting is an optimal-stopping decision. Let $V_t = \max_{d \neq \textsc{wait}} \mathbb{E}[U_d \mid x_{1:t}]$ be the value of the best immediate decision given the audio so far. Listening for a further interval $\Delta t$ costs $\lambda\,\Delta t$ in latency but may raise $V$, for instance by revealing a correction. A one-step look-ahead rule commits as soon as
\begin{equation}
\mathbb{E}\big[V_{t+\Delta t} \mid x_{1:t}\big] - V_t \leq \lambda\,\Delta t,
\label{eq:wait}
\end{equation}
where the expectation is over the audio still to come. This rule favors waiting until the end of the utterance for irreversible actions and allows immediate execution of reversible ones. Section~\ref{sec:sim} tests a simplified, risk-dependent version; estimating the expected future value from partial audio remains open (Section~\ref{sec:challenges}).

\textit{Fuzzy action gating.} Proposition~1 prescribes the ideal decision, but only when $p$ is calibrated and costs are numeric. Fuzzy inference is a practical approximation for three reasons. Its inputs, such as ``the speaker probably matches'' or ``the action is hard to undo,'' are graded and linguistically qualified \cite{zadeh1965fuzzy}. Its rules can be written, read, and audited by designers and domain experts, which matters in regulated applications. And decision-making in a fuzzy environment, where goals and constraints are themselves fuzzy sets, has a well-established foundation \cite{bellman1970decision}. We use type-1 fuzzy sets for clarity; when the inputs are themselves uncertain, as with miscalibrated confidences or noisy addressedness estimates, interval type-2 sets \cite{karnik1999type2} model this second-order uncertainty explicitly, and Section~\ref{sec:sim} evaluates such a gate.

The gate takes the input vector $\mathbf{x} = (\alpha, \sigma, \gamma, \rho, \eta, r)$, holding addressedness, speaker match, content confidence, tool agreement, utterance completeness, and action risk, each normalized to $[0,1]$. Tool agreement is one minus the normalized entropy of the tools' votes on the requested content. If $N \geq 2$ tools return content hypotheses and $n_j$ of them vote for hypothesis $j$,
\begin{equation}
\rho = 1 + \frac{\sum_j p_j \log p_j}{\log N}, \qquad p_j = \frac{n_j}{N},
\label{eq:tool_agreement}
\end{equation}
so $\rho = 1$ when all tools agree and decreases as their votes disperse. Whereas $\rho$ measures only the consistency of tool outputs, content confidence $\gamma$ also weights each tool by its confidence and validity \eqref{eq:fusion}.

Each input is described by linguistic terms such as \emph{Low}, \emph{Medium}, and \emph{High} with membership functions $\mu(\cdot)$. For content confidence, \emph{Low} is 1 below 0.3 and falls linearly to 0 at 0.5, \emph{High} rises from 0 at 0.6 to 1 at 0.8, and \emph{Medium} is a triangular term between them. Rules take the form ``IF $x_1$ is $A_{j1}$ AND $\ldots$ AND $x_n$ is $A_{jn}$ THEN $d = d_j$,'' with $d_j \in \{\textsc{act}, \textsc{wait}, \textsc{ask}, \textsc{silent}\}$. Following Mamdani inference \cite{mamdani1975fuzzy}, a rule fires with the minimum of its antecedent memberships, each decision takes the support of its strongest rule, and the gate selects the decision with the largest support. Ties, including the case in which no rule fires, are broken toward the more conservative decision in the order \textsc{silent}, \textsc{ask}, \textsc{wait}, \textsc{act}, so the gate fails safe. Table~\ref{tab:rules} gives a compact rule base that encodes the principles of Section~\ref{sec:principles}.

\begin{table}[t]
\centering
\caption{Example rule base for the fuzzy action gate. Inputs: addressedness $\alpha$, speaker match $\sigma$, content confidence $\gamma$, tool agreement $\rho$, utterance completeness $\eta$, and action risk $r$. L, M, H: Low, Medium, High; $\neg$L: not Low; blank: any value.}
\label{tab:rules}
\renewcommand{\arraystretch}{1.12}
\footnotesize
\setlength{\tabcolsep}{4pt}
\begin{tabular}{l *{6}{C{0.55cm}} l}
\toprule
& \multicolumn{6}{c}{\textbf{Inputs}} & \\
\cmidrule(lr){2-7}
\textbf{Rule} & $\boldsymbol{\alpha}$ & $\boldsymbol{\sigma}$ & $\boldsymbol{\gamma}$ & $\boldsymbol{\rho}$ & $\boldsymbol{\eta}$ & $\boldsymbol{r}$ & \textbf{Decision} \\
\midrule
$R_1$ & L & & & & & & \textsc{silent} \\
$R_2$ & & L & & & & $\neg$L & \textsc{silent} \\
$R_3$ & & & & & L & & \textsc{wait} \\
$R_4$ & H & H & H & H & H & & \textsc{act} \\
$R_5$ & H & H & M & & H & L & \textsc{act} \\
$R_6$ & H & & M & & H & H & \textsc{ask} \\
$R_7$ & H & & & L & H & & \textsc{ask} \\
$R_8$ & & & L & & H & & \textsc{ask} \\
$R_9$ & M & & $\neg$L & & & & \textsc{ask} \\
$R_{10}$ & H & M & & & & & \textsc{ask} \\
\bottomrule
\end{tabular}
\end{table}

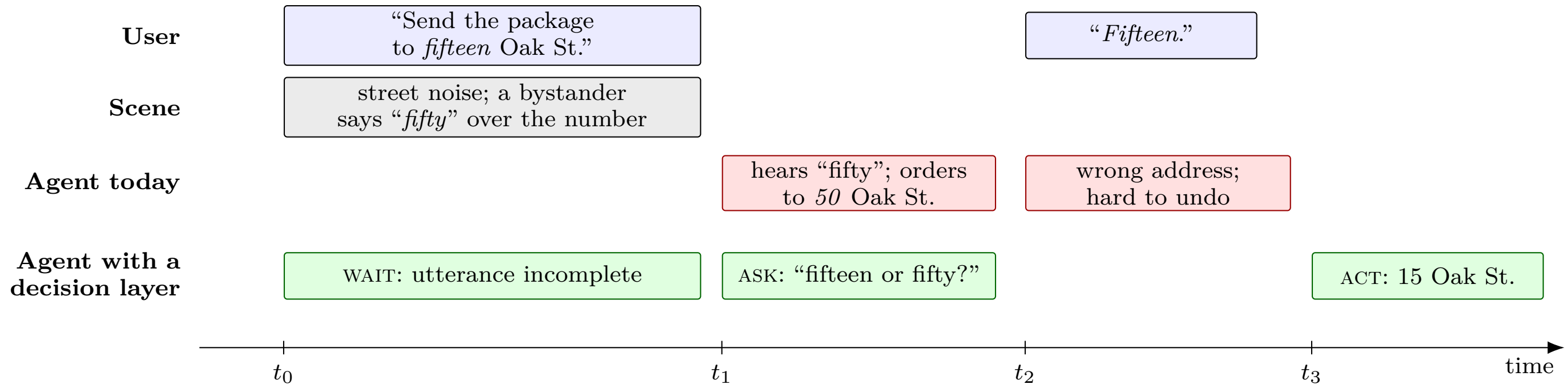
\begin{figure*}[t]
\centering
\resizebox{\textwidth}{!}{%
\begin{tikzpicture}[
  font=\footnotesize,
  seg/.style={draw, rounded corners=1pt, minimum height=5.5mm, inner sep=2pt, align=center},
  user/.style={seg, fill=blue!8},
  noise/.style={seg, fill=black!8},
  bad/.style={seg, fill=red!12, draw=red!60!black},
  good/.style={seg, fill=green!12, draw=green!40!black},
  lab/.style={anchor=east, align=right, font=\footnotesize\bfseries}
]
\draw[-{Latex[length=2mm]}] (0,0) -- (16.2,0) node[below left]{time};
\foreach \x/\t in {1/$t_0$,6.2/$t_1$,9.8/$t_2$,13.2/$t_3$}{\draw (\x,0.08)--(\x,-0.08) node[below]{\t};}
\node[lab] at (-0.1,3.7) {User};
\node[lab] at (-0.1,2.85) {Scene};
\node[lab] at (-0.1,1.95) {Agent today};
\node[lab] at (-0.1,0.85) {Agent with a\\decision layer};
\node[user, text width=4.8cm, anchor=west] at (1,3.7) {``Send the package to \emph{fifteen} Oak St.''};
\node[user, text width=2.6cm, anchor=west] at (9.8,3.7) {``\emph{Fifteen}.''};
\node[noise, text width=4.8cm, anchor=west] at (1,2.85) {street noise; a bystander says ``\emph{fifty}'' over the number};
\node[bad, text width=3.1cm, anchor=west] at (6.2,1.95) {hears ``fifty''; orders to \emph{50} Oak St.};
\node[bad, text width=3.0cm, anchor=west] at (9.8,1.95) {wrong address; hard to undo};
\node[good, text width=4.8cm, anchor=west] at (1,0.85) {\textsc{wait}: utterance incomplete};
\node[good, text width=3.1cm, anchor=west] at (6.2,0.85) {\textsc{ask}: ``fifteen or fifty?''};
\node[good, text width=2.6cm, anchor=west] at (13.2,0.85) {\textsc{act}: 15 Oak St.};
\end{tikzpicture}}
\caption{Anatomy of a decision failure, adapted from \cite{ivry2026voice}: a bystander's ``fifty'' overlaps the user's ``fifteen''. Without a decision layer, the agent commits to the wrong address at $t_1$; with one, it waits, asks at $t_1$, receives the answer at $t_2$, and acts correctly at $t_3$.}
\label{fig:anatomy}
\end{figure*}

\begin{figure}[t]
\centering
\begin{tikzpicture}[
  font=\footnotesize,
  fstep/.style={draw, rounded corners=2pt, fill=black!4, text width=4.4cm, align=center, minimum height=8mm, inner sep=3pt},
  test/.style={draw, rounded corners=2pt, fill=corecolor!6, text width=4.4cm, align=center, minimum height=8mm, inner sep=3pt},
  dec/.style={draw=gatecolor, very thick, rounded corners=2pt, fill=gatecolor!10, text width=4.4cm, align=center, minimum height=8mm, inner sep=3pt},
  note/.style={draw=gatecolor, dashed, rounded corners=2pt, text width=2.2cm, align=center, inner sep=3pt, font=\scriptsize},
  arr/.style={-{Latex[length=1.6mm]}, semithick},
  lbl/.style={font=\scriptsize}
]
\node[fstep] (n1) at (0,0)     {Proposed action: paid order to Oak St.};
\node[test] (n2) at (0,-1.2)  {Utterance complete? ($\eta$ High)};
\node[dec, text width=2.2cm] (w) at (4.1,-1.2) {\textsc{wait} ($R_3$)\\ keep listening};
\node[fstep] (n3) at (0,-2.4)  {$\alpha{=}0.9$,\; $\sigma{=}0.85$,\; $\gamma{=}0.55$,\; $r{=}0.9$};
\node[fstep] (n4) at (0,-3.6)  {Rule support: $R_4, R_5 \approx 0$;\; $R_6 \approx 1$};
\node[dec]  (a)  at (0,-4.8)  {\textsc{ask} ($R_6$): ``fifteen or fifty?''};
\node[note] (chk) at (4.1,-4.8) {Prop.~1 agrees:\\ $p\approx0.42$\\ $< \tau = 0.97$};
\node[fstep] (n5) at (0,-6.0)  {User: ``Fifteen'' $\Rightarrow$ $\gamma$, $\rho$ High};
\node[dec]  (ac) at (0,-7.2)  {\textsc{act} ($R_4$): order to 15 Oak St.};
\draw[arr] (n1) -- (n2);
\draw[arr] (n2) -- node[lbl, above]{no} (w);
\draw[arr] (n2) -- node[lbl, right]{yes} (n3);
\draw[arr] (n3) -- (n4);
\draw[arr] (n4) -- (a);
\draw[gatecolor, dashed] (a) -- (chk);
\draw[arr] (a) -- (n5);
\draw[arr] (n5) -- (ac);
\end{tikzpicture}
\caption{The fuzzy gate on the package order of Fig.~\ref{fig:anatomy} (illustrative values). The gate waits while the utterance is incomplete, asks when content confidence is too low for a high-risk action, and acts once the user's answer resolves the ambiguity.}
\label{fig:gateflow}
\end{figure}
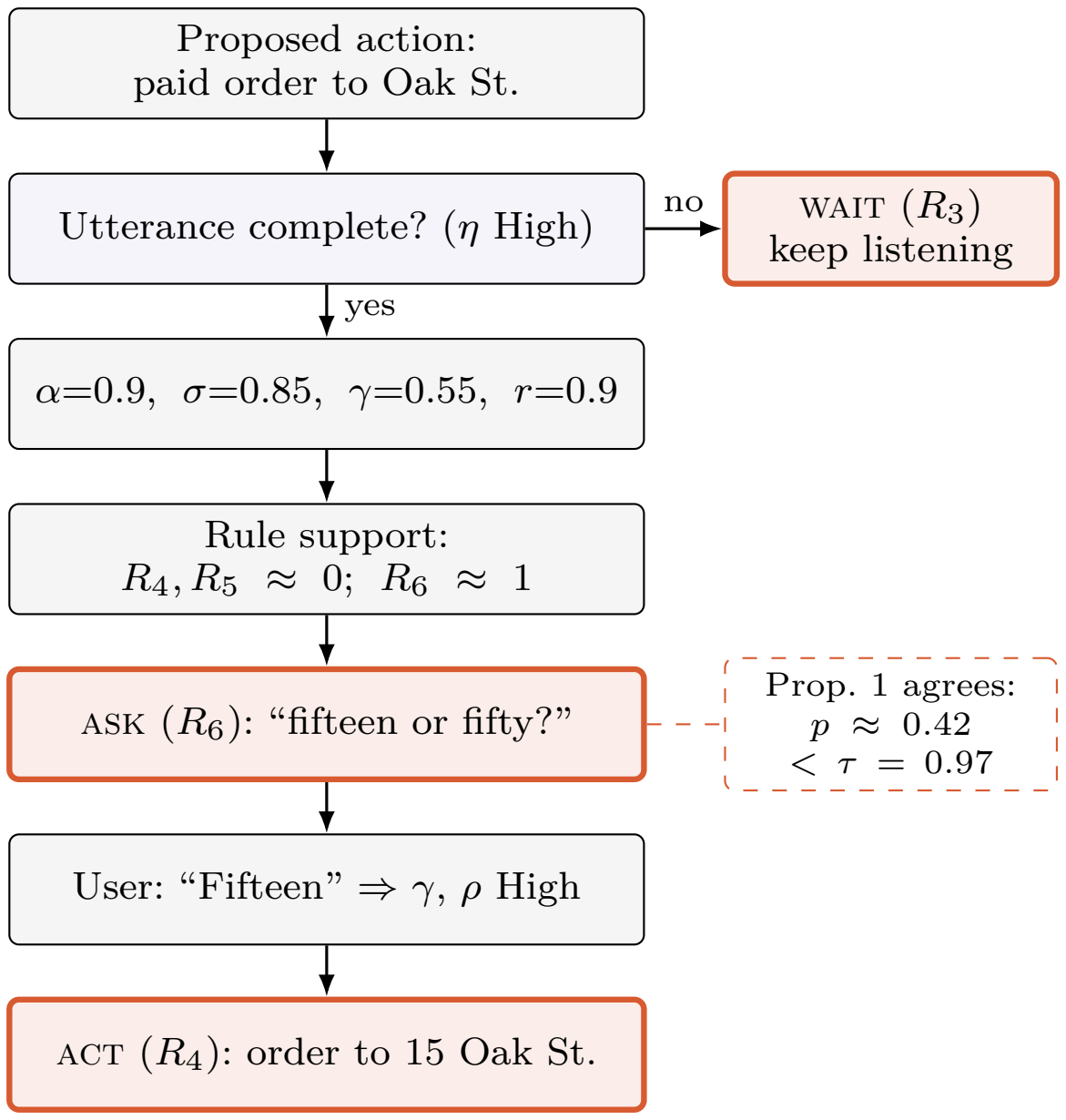

Fig.~\ref{fig:gateflow} traces the gate through the package order of Fig.~\ref{fig:anatomy}, with illustrative rather than measured values. While the user is still speaking, completeness is Low, so $R_3$ fires and the agent waits. Once the utterance ends, addressedness is High ($\alpha = 0.9$) and the speaker matches the account holder ($\sigma = 0.85$), but the overlapped house number has confidence only $\gamma = 0.55$, and the paid, hard-to-reverse order has risk $r = 0.9$. $R_6$ fires with strength close to 1 while $R_4$ and $R_5$ do not, so the agent asks ``fifteen or fifty?'' Proposition~1 agrees: by \eqref{eq:decomp}, $p \approx 0.9 \times 0.85 \times 0.55 \approx 0.42$, far below this action's threshold of $0.97$. Once the user confirms ``fifteen,'' content confidence and tool agreement become High, $R_4$ fires, and the agent acts. A low-risk request, such as reading the weather, would instead proceed at moderate confidence through $R_5$.

\textit{Evolutionary tuning.} Choosing tools, their order, and their thresholds is a combinatorial design problem with conflicting objectives. For a tool chain and its gate parameters $\pi$, we seek the Pareto set of
\begin{equation}
\min_{\pi}\ \mathbf{F}(\pi) = \big(f_{\mathrm{err}}(\pi),\ f_{\mathrm{lat}}(\pi),\ f_{\mathrm{wrong}}(\pi),\ f_{\mathrm{ask}}(\pi)\big),
\end{equation}
namely task error, latency, wrong-action rate, and clarification rate. Multi-objective evolutionary algorithms such as NSGA-II \cite{deb2002nsga2} return a set of trade-offs rather than one solution, so a designer can pick a conservative configuration for emergency response and a fast one for media control. Unlike prompt and workflow search for text agents \cite{guo2024evoprompt, zhang2025aflow}, this search must also weigh real-time latency and the cost of acting on misheard speech.The gate's membership functions form a small, continuous parameter vector well suited to evolutionary search, which Section~\ref{sec:sim} tunes for error cost and ask rate.

\textit{Evidence fusion.} Reliability-weighted fusion combines the outputs of several tools into a single content hypothesis $\hat{y}$ and its confidence $\gamma$. With weights $w_k = v_k c_k$ from the evidence contract \eqref{eq:contract},
\begin{equation}
\hat{y} = \arg\max_{y} \sum_k w_k\,\mathbbm{1}[y_k = y], \qquad \gamma = \frac{\sum_k w_k\,\mathbbm{1}[y_k = \hat{y}]}{\sum_k w_k},
\label{eq:fusion}
\end{equation}
so each tool contributes in proportion to its confidence and the validity of the current conditions. The resulting $\gamma$ is reliability-weighted support for $\hat{y}$ rather than a calibrated probability of correctness, and it presupposes calibrated tool confidences \cite{guo2017calibration}. Validity degrees are themselves natural fuzzy quantities, such as ``valid to the degree that the signal-to-noise ratio is High and overlap is Low.''
\section{Experimental Evaluation}\label{sec:evidence}

If the architecture of Fig.~\ref{fig:arch} is sound, each of its elements should measurably improve decisions. We test claims C1--C7 (Section~\ref{sec:principles}) in two complementary experiments. A controlled simulation (Section~\ref{sec:sim}) isolates the decision rule and tests the claims that concern it alone; recorded audio from two public benchmarks (Section~\ref{sec:real}) tests the claims that need a real model or real audio; and a leave-one-input-out ablation (Section~\ref{sec:buys}) isolates each input. Because our simulator generates evidence from our own modeling assumptions, it can show how large an effect would be if those assumptions hold, but not that the effect occurs in practice. We therefore call a claim \emph{supported} only when recorded audio confirms it. Table~\ref{tab:claims} summarizes the outcome for each claim; the rest of this section gives the details.

\begin{table}[t]
\centering
\caption{Summary of results. Sim: controlled simulation; Rec: recorded audio. A claim is \emph{supported} only when recorded audio confirms it.}
\label{tab:claims}
\renewcommand{\arraystretch}{1.12}
\footnotesize
\setlength{\tabcolsep}{3pt}
\begin{tabularx}{\columnwidth}{P{2.1cm} L P{1.1cm}}
\toprule
\textbf{Claim} & \textbf{Key result} & \textbf{Status} \\
\midrule
C1: a decision layer beats the model acting directly & Rec: side talk executed $41.2 \to 7.9\%$; silence under speaker switch $3.1 \to 65.8\%$ & Supported \\
C2: addressedness and speaker evidence matter most & Sim: wrong actions $14.0 \to 1.4\%$. Rec: side talk executed $36.5 \to 7.4$--$9.8\%$ & Supported \\
C3: risk pricing keeps costly errors rare & Sim: wrong high-risk actions $1.4 \to {\leq}0.05\%$, ${\leq}0.3\%$ under miscalibration. Rec: $5\times$ more without the risk input & Supported \\
C4: addressedness is better judged from audio & Rec: side talk executed $21.6 \to 7.9\%$; silence under speaker switch $24.9 \to 65.8\%$ & Supported \\
C5: reliability-weighted fusion helps & Sim: content accuracy $83.5 \to 88.9\%$; gain shrinks with poor validity estimates & Sim only \\
C6: late commitment avoids acting on corrections & Sim: acting on corrected words $38.6 \to 2.4\%$, at $+0.44$~s & Sim only \\
C7: well-tuned gates perform alike & Sim: utilities within 0.01. Rec: within confidence intervals ($p=0.61$). Fuzzy gates most robust & Supported \\
\bottomrule
\end{tabularx}
\end{table}

\subsection{Experimental Design}\label{sec:design}

The controlled simulation isolates the decision rule: every policy receives identical evidence on identical episodes, so that differences are attributable to the rule alone, and the conditions that generate the evidence can be varied and deliberately violated. The experiment on recorded audio then replaces simulated evidence with the outputs of real models on two public benchmarks. Table~\ref{tab:settings} summarizes both experiments, and Table~\ref{tab:policies} defines the decision policies they compare. In the main simulation all utterances are complete and all tools agree, so rules $R_3$ and $R_7$ are inactive; two extensions activate them.

Four design choices deserve comment. First, the extensions are built to be fair to the competing policies, if anything handicapping our own gate: early self-corrections can be caught even by a rule that commits immediately, some requests contain hesitations for which waiting brings no benefit, and because validity weighting would win by construction if validity were computed from the very conditions that cause failures, we also test a \emph{misspecified} estimator that ignores overlap. Second, the test of C4 keeps the model fixed and changes only its input: the same LALM judges addressedness once from the audio and once from the Whisper transcript. Third, our WearVox figures are not directly comparable with those of the benchmark's authors, who report side-talk rejection as binary accuracy under their own prompt and beamformed audio \cite{lin2026wearvox}; we report the share of side-talk recordings on which a tool call is executed. Fourth, C3 is tested on the valid WearVox queries, since side talk is the same whatever the risk of the requested tool; a high-risk request counts as executed wrongly if the gate acts on content that differs from the reference, such as a misrecognized number or contact name.

We report three kinds of metric throughout. \emph{Errors} are wrong actions; on recorded audio, they are measured as the share of side-talk recordings on which a tool call is executed (lower is better) and as correct silence under speaker switch (higher is better). \emph{Service} is the share of valid requests served (higher is better). \emph{Burden} is the ask rate, the share of episodes in which the agent asks a question. Utility combines all three with the costs of Table~\ref{tab:settings}. Recorded benchmarks contain no answer to a clarifying question, so we score \textsc{ask} decisions under explicit assumptions: a question on a valid request is resolved with probability $q=1$ (optimistic) or $q=0.8$ (realistic), and a question on side talk counts as an interruption of a bystander. Systems are compared with McNemar tests on paired episodes, with a Holm correction across all twelve paired comparisons.

\begin{table*}[t]
\centering
\caption{Experimental settings. All experiments use $G=1$, $\kappa=0.3$, and $M=0.5$.}
\label{tab:settings}
\renewcommand{\arraystretch}{1.12}
\footnotesize
\begin{tabularx}{\textwidth}{P{1.75cm} L L L L}
\toprule
& \textbf{Main simulation} & \textbf{Waiting extension (C6)} & \textbf{Fusion extension (C5)} & \textbf{Recorded audio} \\
\midrule
Episode & One proposed action & One addressed request, revealed in five chunks of 0.4~s & One proposed action whose content comes from three tools & One recording, with the tool call it requests \\
Data & Independent and correlated worlds, each with 200{,}000 training and 200{,}000 test episodes & 200{,}000 test episodes & Correlated world & WearVox side-talk split (582 side-talk, 500 valid queries) \cite{lin2026wearvox}; VGBench speaker switch \cite{zhang2026vgbench} \\
Ground truth & Addressed with probability 0.7; 90\% of speakers authorized; content correct with probability $1/(1+e^{-(\mathrm{SNR}-5)/4})$, SNR uniform in $[-5, 25]$~dB. Correlated world: overlap in 30\% of episodes makes a bystander more likely and lowers content accuracy & Self-correction in 15\% of requests, at a chunk drawn uniformly from the second to the last; content-preserving hesitation in a further 10\% & A recognizer that degrades below 5~dB, a noise-robust recognizer that fails under overlap, and an LALM of moderate accuracy throughout & Benchmark labels \\
Evidence & Noisy $\alpha$, $\sigma$, $\gamma$ from overlapping beta distributions, plus the risk class; in the correlated world, a shared nuisance term depresses $\sigma$ and $\gamma$ & As in the main simulation, plus a noisy completeness score $\eta$ & Calibrated $c_k$ and validity $v_k$ per tool, from noisy estimates of SNR and overlap; agreement $\rho$ from the votes & $\alpha$: Qwen2-Audio's probability of ``yes'' to ``Is this speech directed at the assistant?''; $\gamma$: mean token probability of Whisper large-v3; $\sigma$: ECAPA-TDNN similarity to the enrolled user (214 VGBench items), otherwise 1 \\
Risk and costs & 40\% high-risk ($C=10$), otherwise $C=1$ & As in the main simulation, plus $\lambda=0.05$ per chunk of delay & As in the main simulation & Tools labeled high-risk (payments, messages, device control) or low-risk (information) by a fixed lookup table \\
Tuning and testing & NSGA-II \cite{deb2002nsga2} on 30{,}000 training episodes, population 40, 50 generations, ten seeds; reported at the ask rate of the calibrated product gate & Tuned fuzzy gate of the main simulation & Tuned fuzzy gate of the main simulation & Calibration and NSGA-II on a random 30\% of each benchmark, testing on the remaining 70\%; 1{,}000 bootstrap resamples \\
Compared & Policies of Table~\ref{tab:policies} & Commit immediately; always wait; risk-dependent commit, which waits for completion before high-risk actions only, a simplified form of (\ref{eq:wait}); fuzzy gate with $R_3$ active & Single best tool; majority vote; confidence weighting ($w_k=c_k$); reliability weighting by (\ref{eq:fusion}), correct and misspecified & Policies of Table~\ref{tab:policies}, including the recorded-audio baselines \\
\bottomrule
\end{tabularx}
\end{table*}

\begin{table*}[t]
\centering
\caption{Decision policies compared. Inputs: the evidence each policy uses ($\alpha$: addressedness, $\sigma$: speaker match, $\gamma$: content confidence, $r$: action risk). Sim, Rec: used in simulation and on recorded audio. \pmark: only when placed behind the tuned fuzzy gate.}
\label{tab:policies}
\renewcommand{\arraystretch}{1.1}
\footnotesize
\setlength{\tabcolsep}{4pt}
\begin{tabularx}{\textwidth}{P{3.1cm} L *{4}{C{0.4cm}} C{0.6cm} C{0.6cm}}
\toprule
& & \multicolumn{4}{c}{\textbf{Inputs}} & \multicolumn{2}{c}{\textbf{Used}} \\
\cmidrule(lr){3-6}\cmidrule(lr){7-8}
\textbf{Policy} & \textbf{Rule} & $\alpha$ & $\sigma$ & $\gamma$ & $r$ & \textbf{Sim} & \textbf{Rec} \\
\midrule
Always act & Executes every proposed action &  &  &  &  & \cmark &  \\
\midrule
\multicolumn{8}{l}{\textit{Single signal}} \\
Content threshold & Act if $\gamma \geq 0.8$, else ask; the classical confirmation policy of spoken dialogue systems \cite{bohus2005sorry} &  &  & \cmark &  & \cmark & \cmark \\
\midrule
\multicolumn{8}{l}{\textit{Risk-blind}} \\
Evidence threshold & Act if $\alpha, \sigma > 0.6$ and $\gamma > 0.8$, else ask & \cmark & \cmark & \cmark &  & \cmark & \cmark \\
Calibrated gate & Proposition~1 with the low-risk threshold for every action & \cmark & \cmark & \cmark &  & \cmark &  \\
\midrule
\multicolumn{8}{l}{\textit{Risk-aware}} \\
Always confirm high-risk & Evidence threshold, with every high-risk action turned into a question & \cmark & \cmark & \cmark & \cmark & \cmark & \cmark \\
Calibrated product gate & Proposition~1 on $\alpha$, $\sigma$, $\gamma$ calibrated with 20-bin histograms and combined by (\ref{eq:decomp}); silent if calibrated $\alpha < 0.5$. A one-step, belief-based policy in the spirit of POMDP dialogue managers \cite{williams2007pomdp} & \cmark & \cmark & \cmark & \cmark & \cmark & \cmark \\
Hand-written fuzzy gate & Rules of Table~\ref{tab:rules}, membership functions of Section~\ref{sec:ci} & \cmark & \cmark & \cmark & \cmark & \cmark &  \\
Tuned threshold gate & Separate $\alpha$, $\sigma$, $\gamma$ thresholds per risk class (6 parameters) & \cmark & \cmark & \cmark & \cmark & \cmark & \cmark \\
Tuned fuzzy gate & Fuzzy gate with 12 tuned membership breakpoints & \cmark & \cmark & \cmark & \cmark & \cmark & \cmark \\
Tuned type-2 fuzzy gate & Interval type-2 version with 24 breakpoints; decides from the midpoint of each support interval & \cmark & \cmark & \cmark & \cmark & \cmark &  \\
\midrule
\multicolumn{8}{l}{\textit{Recorded-audio baselines}} \\
Model acts directly & The base agent's own decision to call a tool or remain silent &  &  &  &  &  & \cmark \\
Prompted model & Prompted to remain silent unless addressed by its user; variants add five worked examples (few-shot) or reasoning about who is speaking to whom &  &  &  &  &  & \cmark \\
Detector $\alpha$ & Tuned fuzzy gate fed by a logistic-regression addressedness detector on ECAPA-TDNN and Whisper encoder embeddings, trained on the WearVox tuning split, optionally with inter-microphone level and phase differences & \cmark & \cmark & \cmark & \cmark &  & \cmark \\
Other base models & SALMONN-13B \cite{tang2023salmonn} and Qwen2.5-Omni-7B \cite{xu2025qwen25omni}, acting directly or behind the tuned fuzzy gate & \pmark & \pmark & \pmark & \pmark &  & \cmark \\
\bottomrule
\end{tabularx}
\end{table*}

\subsection{Insights from Simulation}\label{sec:sim}

In short, the simulation shows that evidence about who spoke matters most, that risk-awareness is what keeps costly errors rare, and that the choice of algorithm matters less than its inputs. Table~\ref{tab:sim} reports the main simulation and Fig.~\ref{fig:robust} its robustness when assumptions are violated; Tables~\ref{tab:simext} and~\ref{tab:sens} in the Appendix report the two extensions and a sensitivity analysis. Five findings stand out.

\begin{table*}[t]
\centering
\caption{Controlled simulation, independent world, 200{,}000 held-out episodes. Rates in percent; utility is the mean per episode (higher is better). Tuned gates: mean $\pm$ standard deviation over ten NSGA-II seeds. With 200{,}000 episodes, 95\% confidence intervals of the rates are below $\pm 0.2$ points.}
\label{tab:sim}
\renewcommand{\arraystretch}{1.1}
\footnotesize
\begin{tabularx}{\textwidth}{P{2.3cm} L P{1.35cm} P{1.45cm} P{1.45cm} P{1.2cm} P{1.45cm} P{1.3cm}}
\toprule
\textbf{Group} & \textbf{Policy} & \textbf{Wrong actions} & \textbf{High-risk wrong} & \textbf{Side talk executed} & \textbf{Ask rate} & \textbf{Requests served} & \textbf{Utility} \\
\midrule
No gate & Always act & 58.7 & 58.5 & 100.0 & 0.0 & 59.1 & $-2.28$ \\
Single signal & Content threshold & 14.0 & 14.2 & 34.6 & 65.5 & 94.8 & $-0.19$ \\
\midrule
\multirow{2}{*}{Risk-blind} & Evidence threshold & 1.4 & 1.4 & 1.2 & 58.9 & 96.6 & 0.43 \\
 & Calibrated gate & 1.3 & 1.4 & 1.0 & 55.7 & 95.0 & 0.42 \\
\midrule
\multirow{6}{*}{Risk-aware} & Always confirm high-risk & 0.8 & 0.00 & 0.7 & 65.6 & 97.2 & \textbf{0.47} \\
 & Calibrated product gate & 0.8 & 0.01 & 0.6 & 62.0 & 95.6 & 0.46 \\
 & Hand-written fuzzy gate & 5.4 & 1.9 & 0.6 & 50.3 & 87.9 & 0.32 \\
 & Tuned threshold gate & $1.6 \pm 0.3$ & $0.05 \pm 0.07$ & $1.4 \pm 0.8$ & $62.1 \pm 1.0$ & $96.5 \pm 0.3$ & $0.46 \pm 0.00$ \\
 & \textbf{Tuned fuzzy gate} & $1.1 \pm 0.3$ & $0.02 \pm 0.02$ & $0.2 \pm 0.2$ & $62.0 \pm 0.7$ & $94.5 \pm 1.1$ & $0.45 \pm 0.01$ \\
 & Tuned type-2 fuzzy gate & $1.0 \pm 0.3$ & $0.02 \pm 0.02$ & $0.2 \pm 0.1$ & $61.8 \pm 0.8$ & $94.6 \pm 1.0$ & $0.45 \pm 0.01$ \\
\bottomrule
\end{tabularx}
\end{table*}

\begin{figure*}[t]
\centering
\includegraphics[width=\textwidth]{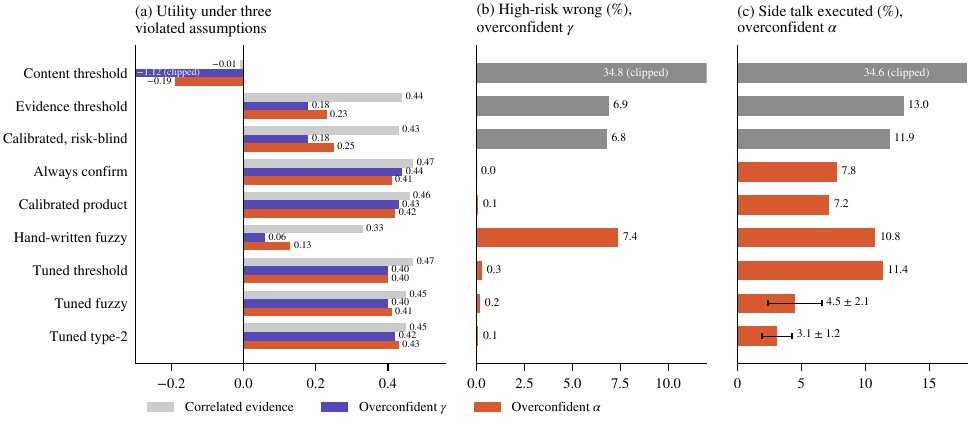}
\caption{Robustness without re-tuning. (a) Utility under correlated evidence and under overconfident content confidence ($\gamma \mapsto \gamma^{0.4}$) or addressedness ($\alpha \mapsto \alpha^{0.4}$). (b) Wrong high-risk actions with overconfident $\gamma$. (c) Executed side talk with overconfident $\alpha$ (mean $\pm$ standard deviation over ten seeds). Orange: risk-aware gates; gray: risk-blind or single-signal policies.}
\label{fig:robust}
\end{figure*}

\begin{figure}[t]
\centering
\includegraphics[width=\columnwidth]{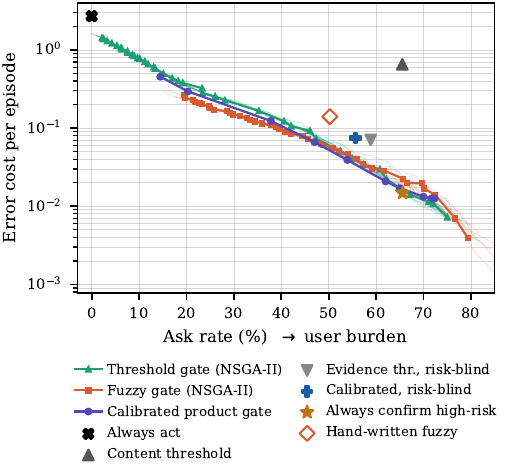}
\caption{Error cost versus ask rate on held-out episodes of the independent world (log scale). Thin lines are the NSGA-II fronts of ten seeds for each tuned gate; marked lines are the seeds with median hypervolume. The calibrated product curve sweeps the ask cost $\kappa$. Baselines are single operating points.}
\label{fig:pareto}
\end{figure}

\begin{enumerate}
\item \emph{Evidence about who spoke and to whom matters most (C2).} The content-threshold policy asks almost as often as the best gates, yet executes 34.6\% of commands not addressed to the agent. Adding addressedness and speaker evidence, even with fixed heuristic thresholds, cuts wrong actions from 14.0\% to 1.4\%, as the evidence bound of Section~\ref{sec:theory} predicts.

\item \emph{Risk-awareness keeps costly errors rare, even under miscalibration (C3).} Risk-blind gates execute wrong high-risk actions at about 1.4\%, 30 to 70 times more often than tuned risk-aware gates. With two sharply separated risk classes, Proposition~1 reduces to confirming every high-risk action, which is the best policy here (utility 0.47) and which the calibrated product gate implements by acting on fewer than 1\% of high-risk requests. With an overconfident recognizer, risk-blind gates fall to a utility of 0.18 with about 7\% wrong high-risk actions, whereas every calibrated or tuned risk-aware gate stays at or below 0.3\% (Fig.~\ref{fig:robust}); the policy ranking is also unchanged in the correlated world. Which risk-aware gate is best depends on the cost of asking and the risk mix: when high-risk actions are 10--20\% of requests or $\kappa = 1$, the tuned gates are best, and the ask rate of the tuned fuzzy gate falls from 62\% to 35--45\% (Table~\ref{tab:sens}).

\item \emph{Inputs matter more than the algorithm, and fuzzy gates are the most robust (C7).} At matched ask rates, the tuned gates and the calibrated product gate differ in utility by at most 0.01, and their fronts overlap between 30\% and 70\% ask rates (Fig.~\ref{fig:pareto}). The tuned threshold gate reaches the largest hypervolume ($0.779 \pm 0.001$, against $0.754 \pm 0.021$ for the tuned fuzzy gate and $0.753$ for the calibrated product curve), mainly because it also reaches very low ask rates. Fuzzy gates differ in robustness: with overconfident addressedness, the tuned fuzzy gate executes $4.5 \pm 2.1$\% of side talk, against $7.2$--$13.0$\% for the non-fuzzy gates that use all three evidence terms, because its rules route intermediate addressedness to a question. The interval type-2 gate lowers this to $3.1 \pm 1.2$\%, a difference from type-1 that is not significant (paired Wilcoxon signed-rank test over ten seeds, $p = 0.084$). Two cautions apply. The untuned hand-written fuzzy gate performs worst of all gates (utility 0.32; 7.4\% wrong high-risk actions under an overconfident recognizer), so expert rules must be tuned on data. And a single-objective evolution strategy that maximized expected utility converged to a gate that never acts and asks 78\% of the time, so user burden must be optimized as a separate objective.

\item \emph{Committing late avoids acting on corrections at little cost (C6).} Committing immediately acts on corrected words in 38.6\% of high-risk requests that contain a self-correction; risk-dependent commitment reduces this to 2.4\% while adding 0.44~s to the mean commit delay (Table~\ref{tab:simext}). Always waiting removes the same errors but delays every action by 1.08~s and gains nothing on hesitations, and the fuzzy gate with $R_3$ active comes close to the best rule (utility 0.37 against 0.39).

\item \emph{Tools that report their validity yield better evidence, if validity is estimated well (C5).} Under frequent overlap, reliability weighting raises the accuracy of the fused content from 83.5\% with confidence weighting to 88.9\% and lowers wrong actions from 2.6\% to 1.7\% (Table~\ref{tab:simext}). With the misspecified validity estimator, the gains shrink to 85.6\% and 2.3\%.
\end{enumerate}

\subsection{Validation on Recorded Audio}\label{sec:real}

\begin{table}[t]
\centering
\caption{Results on recorded audio (held-out 70\%), with Qwen2-Audio as the base agent unless stated otherwise. WearVox: side talk executed (lower is better), valid requests served, and ask rate. VGBench: correct silence under speaker switch (higher is better). Bold: the tuned fuzzy gate, the representative gate discussed in the text. Spatial cues are available on WearVox only. 95\% bootstrap confidence intervals are at most $\pm 2.3$ points on WearVox and $\pm 2.9$ points on VGBench.}
\label{tab:real}
\renewcommand{\arraystretch}{1.1}
\footnotesize
\setlength{\tabcolsep}{3.5pt}
\begin{tabularx}{\columnwidth}{L P{0.95cm} P{1.1cm} P{0.65cm} P{1.1cm}}
\toprule
& \multicolumn{3}{c}{\textbf{WearVox}} & \textbf{VGBench} \\
\cmidrule(lr){2-4}\cmidrule(lr){5-5}
\textbf{System} & \textbf{Side talk exec.} & \textbf{Served} & \textbf{Asks} & \textbf{Correct silence} \\
\midrule
\multicolumn{5}{l}{\textit{Decision policies}} \\
Model acts directly & 41.2 & 88.4 & 0.0 & 3.1 \\
Content threshold & 36.5 & 79.0 & 24.7 & 6.8 \\
Evidence thr., risk-blind & 9.8 & 83.1 & 31.5 & 61.2 \\
Always confirm high-risk & 9.8 & 82.6 & 38.9 & 61.2 \\
Tuned threshold gate & 7.4 & 85.0 & 33.0 & 66.5 \\
\textbf{Tuned fuzzy gate} & \textbf{7.9} & \textbf{84.6} & \textbf{33.4} & \textbf{65.8} \\
Calibrated product gate & 8.6 & 84.2 & 34.1 & 63.9 \\
\midrule
\multicolumn{5}{l}{\textit{Addressedness from the transcript (C4)}} \\
Evidence thr., transcript $\alpha$ & 24.3 & 81.7 & 29.8 & 22.5 \\
Tuned fuzzy, transcript $\alpha$ & 21.6 & 83.0 & 31.2 & 24.9 \\
\midrule
\multicolumn{5}{l}{\textit{Prompting baselines}} \\
Prompted model & 27.3 & 86.1 & 0.0 & 14.7 \\
Prompted, few-shot & 22.6 & 85.4 & 0.0 & 19.8 \\
Prompted, reasoning (best) & 19.4 & 84.7 & 0.0 & 23.5 \\
\midrule
\multicolumn{5}{l}{\textit{Independent addressedness detector}} \\
Tuned fuzzy, detector $\alpha$ & 6.2 & 85.3 & 31.8 & 58.4 \\
Tuned fuzzy, detector $\alpha$ + spatial & 4.1 & 86.2 & 29.5 & -- \\
Tuned fuzzy, mean of agent and detector $\alpha$ & 5.3 & 85.6 & 30.9 & 66.9 \\
\midrule
\multicolumn{5}{l}{\textit{Other base models}} \\
SALMONN, acts directly & 46.8 & 85.9 & 0.0 & 2.4 \\
SALMONN, tuned fuzzy gate & 9.1 & 83.8 & 34.6 & 61.7 \\
Qwen2.5-Omni, acts directly & 33.9 & 87.2 & 0.0 & 9.6 \\
Qwen2.5-Omni, tuned fuzzy gate & 7.1 & 85.1 & 32.2 & 64.4 \\
\bottomrule
\end{tabularx}
\end{table}

\begin{figure}[t]
\centering
\includegraphics[width=\columnwidth]{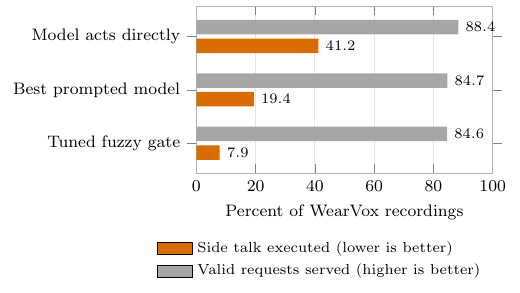}
\caption{Main result on recorded audio (WearVox, held-out 70\%, Qwen2-Audio as base agent). The decision layer executes a fifth as much side talk as the model acting directly, and less than half as much as the best prompted model, while serving almost as many valid requests (served at $q=1$; 79.0\% for the gate at $q=0.8$). Table~\ref{tab:real} gives all systems.}
\label{fig:headline}
\end{figure}

On recorded audio, an explicit gate cuts executed side talk from 41\% to 7--10\%, outperforms prompting, and does so across base models; its main cost is questions (Fig.~\ref{fig:headline}). The text compares three representative systems: the model acting directly, the best prompted model, which reasons about who is speaking to whom, and the tuned fuzzy gate. Table~\ref{tab:real} reports all systems.

\textit{Gate versus model (C1, C2, C7).} Left to decide on its own, the model executes 41.2\% of WearVox side talk and remains silent in only 3.1\% of VGBench speaker switches. The tuned fuzzy gate cuts executed side talk to 7.9\% and raises correct silence to 65.8\%, while serving 84.6\% of valid requests against 88.4\%. Addressedness is what makes the difference: a content threshold barely helps (36.5\%), whereas every gate that uses the addressedness estimate executes 7.4--9.8\% (C2). Every gate differs from the model acting directly ($p < 0.001$), whereas the tuned threshold and fuzzy gates do not differ from each other ($p = 0.61$; C7). Remaining errors come mostly from short utterances and from nearby speakers facing the wearer, where a single-channel estimate is weakest.

\textit{Audio versus transcript (C4).} When the same model judges addressedness from the Whisper transcript instead of the audio, the tuned fuzzy gate executes 21.6\% of side talk instead of 7.9\%, and correct silence under speaker switch falls from 65.8\% to 24.9\% ($p_{\mathrm{Holm}} < 0.001$), because side talk often uses the same words as a request.

\textit{Gate versus prompting.} The best prompted model still executes 19.4\% of side talk, more than twice the gate's 7.9\% at comparable service, and simpler prompts do worse (22.6--27.3\%). Prompting changes how the model judges addressedness but gives it no way to weigh that judgment against the risk of the action, or to ask rather than act. The result holds across base models: the gate cuts executed side talk from 46.8\% to 9.1\% for SALMONN and from 33.9\% to 7.1\% for Qwen2.5-Omni. An independent addressedness detector does better still on WearVox (6.2\%, or 4.1\% with multichannel spatial cues) but generalizes less well to VGBench (58.4\% correct silence, against 65.8\%); averaging it with the agent's own estimate combines their strengths (5.3\% side talk, 66.9\% correct silence), as Section~\ref{sec:alt} discusses.

\begin{figure*}[t]
\centering
\includegraphics[width=\textwidth]{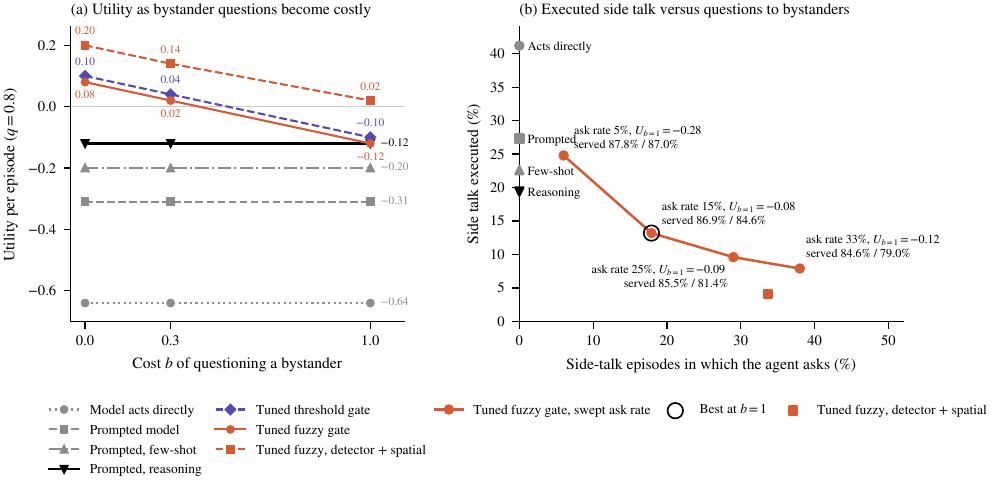}
\caption{Cost of questioning bystanders on WearVox (held-out 70\%, $q=0.8$; costs as in the simulation, mean wrong-action cost 4.6). (a) Utility as the cost $b$ of each question to a bystander grows. (b) Executed side talk against questions to bystanders as the ask rate of the tuned fuzzy gate is swept; labels give the ask rate, the utility at $b=1$, and valid requests served at $q=1$ / $q=0.8$. Systems that never ask lie on the vertical axis.}
\label{fig:bystander}
\end{figure*}

\textit{Cost of questions.} The gates ask in 24.6--28.0\% of valid requests and in 33.7--38.0\% of side-talk episodes. With realistic clarification ($q = 0.8$), the tuned fuzzy gate therefore serves 79.0\% of valid requests (81.3\% with the detector and spatial cues), against 84.7\% for the best prompted model, which executes 2.5 times as much side talk. Whether this trade-off pays depends on the cost $b$ of each question to a bystander (Fig.~\ref{fig:bystander}a). For $b \le 0.3$, every gate beats every prompted model by a wide margin. At $b = 1$, the cost of a wrong low-risk action, the tuned fuzzy gate only ties the best prompted model, while the detector-fed gate still wins. Lowering the ask rate restores the margin (Fig.~\ref{fig:bystander}b): at $b = 1$, the best operating point asks 15\% of the time instead of 33\%, executes 13.2\% of side talk, a third less than the best prompted model, and questions 17.9\% of side-talk episodes. Unlike a prompt, the gate's operating point can thus be tuned to how costly interruptions are in a given deployment.

\subsection{Ablating Each Input}\label{sec:buys}

\begin{figure*}[t]
\centering
\includegraphics[width=\textwidth]{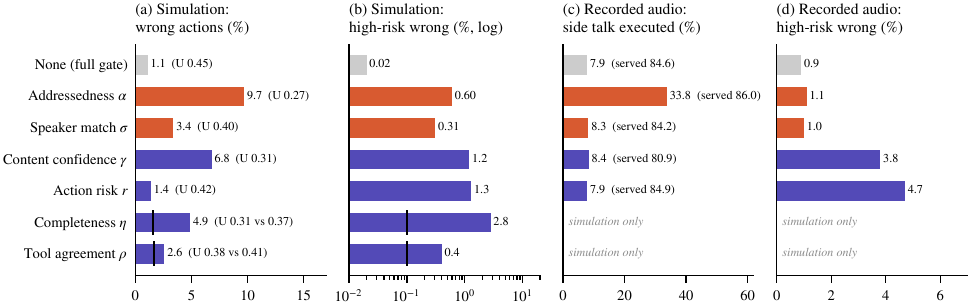}
\caption{Leave-one-input-out ablation: each input of the tuned fuzzy gate is set to a constant and the gate is retuned at the same ask rate. (a, b) Simulation, independent world; completeness and tool agreement use the streaming and fusion episodes, where black ticks mark the full gate. U: utility. (c, d) Recorded WearVox audio; served at $q=1$; high-risk wrong: valid high-risk queries executed with wrong content. Orange: addressedness and speaker evidence.}
\label{fig:ablation}
\end{figure*}

Every input earns its place, but addressedness and speaker evidence earn the most. To show this, we ablate the tuned fuzzy gate itself: each input in turn is set to a constant, and the gate is retuned with NSGA-II at the same ask rate (Fig.~\ref{fig:ablation}). Unlike comparisons between policies, this changes exactly one element of the architecture. Removing addressedness causes the largest loss in both settings: wrong actions in simulation rise almost ninefold, and executed side talk on recorded audio rises to 33.8\%, close to the content threshold. Removing content confidence costs mainly served requests, because misheard commands are no longer questioned.

Removing the risk input leaves overall wrong actions nearly unchanged but multiplies wrong high-risk actions by 65 in simulation and by five on recorded audio (C3). The extra errors on recorded audio are mostly misrecognized numbers and names that the full gate turns into questions; the effect is smaller than in simulation because real recognition errors on high-risk queries are rarer than the simulation assumes. Completeness and tool agreement matter only in the episodes where they vary, but there they matter; such episodes exist only in simulation. Speaker match matters little on WearVox, where enrollment audio is unavailable, but on the 214 VGBench items with enrollment audio, removing it lowers correct silence from 65.8\% to 52.1\%. Because the VGBench switch changes distance rendering and timing as well as the speaker \cite{zhang2026vgbench}, part of this effect may reflect channel rather than identity cues.

\section{Discussion}\label{sec:action}

By the standard set at the start of Section~\ref{sec:evidence}, five claims (C1--C4 and C7) are supported on recorded audio, whereas the evidence contract (C5) and late commitment (C6) are supported in simulation only and remain hypotheses for recorded audio. What the evidence establishes is relative rather than absolute: under identical evidence, the decision layer and its inputs change wrong-action rates by more than an order of magnitude, but the absolute numbers should not be read as predictions for deployed systems (Section~\ref{sec:challenges}). Within these limits, the evidence supports the position and yields four practical implications.

\begin{tcolorbox}[colback=mygray,colframe=black!60,boxrule=0.4pt,arc=1pt,left=4pt,right=4pt,top=3pt,bottom=3pt]
\textbf{Implications for designers.} (1)~Route every consequential tool call through an explicit decision layer rather than letting the model act directly. (2)~Feed that layer addressedness and speaker evidence estimated from the audio, not only transcript confidence. (3)~Price actions by risk: ask or wait before irreversible actions, act promptly on reversible ones. (4)~Tune how often the agent asks to how costly interruptions are in the deployment.
\end{tcolorbox}

The rest of this section addresses the principal objections to the position, proposes criteria for evaluating the decisions of audio agents, and sets out the questions that remain open.

\subsection{Alternative Views}\label{sec:alt}

\textit{``Larger models or end-to-end training will learn to decide.''} Scale alone has not done so: abstention ability is largely independent of task-solving ability in text agents \cite{liu2026agentabstain}, and in our experiments neither prompting nor a more recent base model closes the gap (Section~\ref{sec:real}), as the evidence bound of Section~\ref{sec:theory} predicts. Targeted training is more promising. Supervised training on VGBench mutes 91.3\% of switched commands while still choosing the correct tool for all nearby wearer commands and text-only controls \cite{zhang2026vgbench}, well above the 65.8\% of our training-free gate, and multi-turn reinforcement learning already teaches audio models when to invoke tools \cite{wang2026hiu}. Such training complements the decision layer rather than replacing it. A model trained to reject a known kind of shift does not price the risk of the action it is about to take, cannot choose to ask rather than act, and must be retrained when the deployment, its risks, or its acoustic conditions change. Its judgment is, however, an excellent addressedness estimate $\alpha$, which the layer combines with speaker, content, and risk evidence, exactly as it does with the independent detector of Section~\ref{sec:real}; and its training reward must still price irreversibility and user burden, as Proposition~1 and Fig.~\ref{fig:pareto} make explicit.

\textit{``Cascaded pipelines already solve this.''} A cascade can insert confirmation steps, but it discards the acoustic evidence on which addressedness and speaker decisions depend \cite{chen2026listen}: on recorded audio, judging addressedness from the transcript nearly triples executed side talk (Section~\ref{sec:real}). The decision layer can sit behind a cascade, but only if the cascade passes on more than the transcript.

\textit{``Industry already confirms risky actions.''} It does, and in our simulation confirming every high-risk action is the best policy when risk falls into two clear classes. That policy works, however, only because it is built on addressedness and speaker evidence (Table~\ref{tab:sim}); a confirmation rule driven by transcript confidence would still execute low-risk side talk. The decision layer generalizes confirmation to graded costs, silence, and timing.

\textit{``The gain comes from an existing detector, not the architecture.''} A detector does deliver the best WearVox result, but only because the architecture gives it somewhere to act: current agents have no step at which a device-directed speech detector can stop a tool call. The decision layer accepts any addressedness estimate, combines it with speaker, content, and risk evidence, and allows estimates to be exchanged without retraining the agent. That the best estimate combines the agent's own judgment with a detector is an argument for this modularity, not against it.

\textit{``Asking and waiting will make agents annoying.''} They can, and naive optimization indeed produces an agent that always asks (Section~\ref{sec:sim}). The remedy is to price the choice rather than remove it: (\ref{eq:tau}) asks only when the expected cost of a wrong action exceeds that of a question, and multi-objective design lets each application set its own operating point (Fig.~\ref{fig:bystander}). A question such as ``fifteen or fifty?'' costs seconds; a package sent to the wrong address costs days.

\textit{``Fuzzy logic is outdated.''} We do not propose fuzzy systems as a replacement for neural perception, but as the decision interface, where inputs are graded, costs are linguistic, and rules must be audited. Fuzzy gates do not outperform tuned thresholds on average; their measured advantage is robustness to overconfident addressedness (Section~\ref{sec:sim}), and their rules remain readable and can be refined from logged outcomes \cite{jang1993anfis}.

\subsection{Decision-Level Evaluation Metrics}\label{sec:eval}

Most audio benchmarks measure listening \cite{sakshi2025mmau}; they rarely ask whether an agent took the right action, at the right time, for the right person. Paired protocols point the way: VGBench fixes the words and changes only their acoustic source \cite{zhang2026vgbench}, TRACE compares the same task on clean and stressed recordings \cite{ivry2026voice}, and paired act-or-abstain designs prevent any constant policy from scoring well \cite{liu2026agentabstain}, whereas completion-only benchmarks cannot measure abstention competence \cite{ojewale2026abstention}. We generalize these ideas into decision-level metrics, listed in Table~\ref{tab:eval}.

Three guidelines follow. First, \emph{pair conditions}: compare the same task under clean and stressed audio, or with the same words from different sources, so that differences isolate the decision rather than the content. Second, \emph{score actions, not transcripts}: a correct transcript followed by a wrong action is a failure. Third, \emph{report trade-offs}: an agent that never acts has no wrong actions, so wrong-action rates must be reported together with task completion and clarification rates, ideally as Pareto curves such as Fig.~\ref{fig:pareto}.

\begin{table*}[t]
\centering
\caption{Decision-level evaluation metrics for agentic audio AI. \cmark: decision that the metric tests.}
\label{tab:eval}
\renewcommand{\arraystretch}{1.12}
\footnotesize
\setlength{\tabcolsep}{4pt}
\begin{tabularx}{\textwidth}{P{2.3cm} L C{1.35cm} C{1.05cm} C{0.85cm} C{1.35cm} C{0.95cm} C{0.85cm} P{3.3cm}}
\toprule
& & \multicolumn{6}{c}{\textbf{Decision tested}} & \\
\cmidrule(lr){3-8}
\textbf{Metric} & \textbf{Definition} & \textbf{Whether} & \textbf{Whom} & \textbf{What} & \textbf{Which\newline evidence} & \textbf{When} & \textbf{Safe} & \textbf{Example protocol} \\
\midrule
Wrong-action rate & Fraction of episodes in which an executed action differs from the user's intent & \cmark & \cmark & \cmark & \cmark & \cmark & \cmark & Paired clean and stressed recordings \cite{ivry2026voice} \\
Correct-silence rate & Fraction of non-addressed commands that are not executed & \cmark & \cmark &  &  &  &  & Fixed words with changed speaker or scene \cite{zhang2026vgbench} \\
Clarification precision and rate & Share of \textsc{ask} decisions that were needed, and how often the agent asks &  &  & \cmark & \cmark &  &  & Inject ambiguous numbers and names at controlled noise levels \\
Commit latency & Time from the end of the decisive words to action commitment &  &  &  &  & \cmark &  & Utterances with mid-sentence self-corrections \\
Recovery rate & Fraction of wrong or premature actions that are detected and corrected &  &  & \cmark &  & \cmark &  & Scripted corrections after an action \cite{ivry2026voice} \\
User effort & Extra turns or repetitions needed to complete the task & \cmark & \cmark & \cmark & \cmark & \cmark & \cmark & Simulated or real users under stress conditions \cite{ivry2026voice} \\
Attack success rate & Fraction of adversarial audio instructions that are executed &  &  &  &  &  & \cmark & Narrative and hidden-instruction attacks \cite{yu2026narrative} \\
Calibration error & Gap between reported confidence and observed accuracy, per tool and overall &  &  &  & \cmark &  &  & Reliability diagrams by acoustic condition \cite{guo2017calibration} \\
\bottomrule
\end{tabularx}
\end{table*}

\subsection{Limitations and Future Work}\label{sec:challenges}\label{sec:limitations}

Our evidence has limits, and most of them mark open problems for the field. Two apply throughout. The simulation abstracts perception into scores with assumed distributions and tests only one form of dependence between evidence terms; on recorded audio, all three base models are open-weight LALMs with 7 to 13 billion parameters, and no proprietary model is tested. We discuss the remaining limitations below along with future research opportunities to address them.

\textit{Perception for decisions.} Speaker match is set to 1 where no enrollment audio exists, so our experiments test addressedness more thoroughly than speaker attribution. The independent detector is trained on part of WearVox and may not transfer to other devices. The test of C4 queries one model with and without audio, so it shows that the audio helps this model, not every model. More broadly, addressedness, speaker attribution, and spatial cues matter more for decisions than further gains in word accuracy \cite{lin2026wearvox, mallidi2018devicedirected}; separation, localization, and scene analysis matter most where they supply calibrated evidence to the decision layer.

\textit{Calibration and guarantees.} The decision-theoretic gate needs calibrated confidences, and even a tuned fuzzy gate needs scores whose ranking is stable. LALM calibration across acoustic conditions is largely unmeasured. Selective classification \cite{geifman2017selective} and conformal prediction \cite{ren2023knowno} could bound the rate of wrong actions, once extended to streaming, non-stationary audio.

\textit{Claims tested only in simulation.} Late commitment (C6) and the evidence contract (C5) rest on assumed conditions: utterances arrive in fixed chunks, self-corrections and hesitations follow assumed rates, and tool failure conditions are assumed rather than measured. The waiting rule (\ref{eq:wait}) requires estimating how much the remaining audio is likely to change the decision, and the evidence contract requires tools whose validity is estimated from measured operating ranges. Testing either on recorded audio needs benchmarks that current practice does not provide: recorded speech with natural self-corrections, and logs of when each tool fails.

\textit{Graded costs and many actions.} Our experiments use two risk classes, which favors a simple confirmation rule. We expect tuned gates to matter more when costs are graded, when confirmation itself is costly, or when many action types must be handled; eliciting such costs and testing gates under them remains to be done.

\textit{Asking well.} Questions are scored under assumed resolution rates rather than with real users. How to phrase targeted clarifying questions, how often they are resolved, and how the tolerable ask rate depends on task, culture, and user are open human-factors questions that Proposition~1 prices but does not answer.

\textit{Learning gates from little data.} Logged decision outcomes are scarce and sensitive. Expert-written rules make a fuzzy gate a safe, interpretable starting policy, and neuro-fuzzy methods such as ANFIS \cite{jang1993anfis} can then learn its membership functions and rule weights while the rules stay readable, keeping the gate auditable in regulated settings such as healthcare and public safety. Evolutionary search and safe reinforcement learning can refine it further, for example the timing policy of (\ref{eq:wait}) with rewards that penalize wrong actions, latency, and questions \cite{sutton2018rl}; preference-based alignment \cite{chen2025audio} and multi-turn reinforcement learning \cite{wang2026hiu} are already used to train audio models. Evaluating candidate gates, however, requires realistic simulators, a research problem in its own right.

\textit{Security, privacy, and consent.} Defenses against audio-borne attacks lag behind attacks \cite{yu2026narrative, luo2026lalmsurvey}, and always-listening agents need privacy protection for the people they overhear, a challenge that privacy-preserving audio surveillance frameworks have begun to address \cite{rathore2025towards}. The decision layer raises its own privacy questions. Estimating addressedness and speaker match requires analyzing speech that was not meant for the agent, including the voices of bystanders who never consented. We recommend that these estimates be computed on device, that audio judged not addressed be discarded rather than stored, that speaker profiles be enrolled only with consent, and that the evidence logs used to audit decisions record scores rather than raw audio. Costs also differ by domain: a missed fall in elderly care \cite{doukas2010emergency} and a misrouted order call for different gate parameters, which multi-objective design makes explicit rather than hiding them in a single loss.

\section{Conclusion}\label{sec:conclusion}

Audio AI has become far better at recognizing what was said than at deciding what to do with it. We traced six recurring decision failures to one architectural gap, the absence of an explicit step between reasoning and action. Better perception alone cannot close it: the probability that an action is correct is bounded by the agent's belief about who spoke and to whom, and the evidence required rises with the cost of being wrong. We proposed a reference architecture built around a dedicated decision layer, fed by evidence on addressedness, speaker, content, completeness, and risk, with an evidence contract for tools and computational intelligence methods to build the layer. The evidence supports the proposal. In simulation and on recorded audio alike, the layer's inputs matter far more than the algorithm that implements it, and evidence about the speaker and the addressee matters most. On egocentric recordings, the layer reduces executed side talk from 41\% to 7--10\% while serving most valid requests, a gain that neither better prompting nor newer base models match. An agent that asks ``fifteen or fifty?'' before placing the order is not a weaker listener, but a better decision-maker. We expect progress in agentic audio AI to depend not only on how well agents hear but also on how wisely they decide what to do with what they hear.

\appendices
\section{Supplementary Results}\label{app:supp}
This appendix collects the results of the simulation extensions (Table~\ref{tab:simext}) and of the sensitivity analysis (Table~\ref{tab:sens}), which are discussed in Section~\ref{sec:sim}.

\begin{table}[t]
\centering
\caption{Simulation extensions, held-out episodes. Top: commit rules on streaming episodes with self-corrections at random positions and content-preserving hesitations (C6). Bottom: content fusion with unreliable tools under frequent overlap (C5). WA-corr: wrong high-risk actions caused by self-corrections, as a share of high-risk requests that contain one (\%); Delay: mean time from request onset to commitment, for a 2-s request (s); Acc.: accuracy of the fused content (\%); WA: wrong actions (\%); U: utility.}
\label{tab:simext}
\renewcommand{\arraystretch}{1.1}
\footnotesize
\begin{tabularx}{\columnwidth}{L P{1.0cm} P{1.0cm} P{0.8cm}}
\toprule
\textbf{Commit rule} & \textbf{WA-corr} & \textbf{Delay} & \textbf{U} \\
\midrule
Commit immediately & 38.6 & 0.92 & 0.27 \\
Always wait & 2.4 & 2.00 & 0.34 \\
Risk-dependent commit & 2.4 & 1.36 & 0.39 \\
Fuzzy gate with $R_3$ & 6.1 & 1.81 & 0.37 \\
\midrule
\textbf{Content fusion} & \textbf{Acc.} & \textbf{WA} & \textbf{U} \\
\midrule
Single best tool & 78.4 & 3.9 & 0.36 \\
Majority vote & 81.2 & 3.1 & 0.38 \\
Confidence-weighted & 83.5 & 2.6 & 0.40 \\
Reliability-weighted, misspecified & 85.6 & 2.3 & 0.41 \\
Reliability-weighted & 88.9 & 1.7 & 0.43 \\
\bottomrule
\end{tabularx}
\end{table}

\begin{table}[t]
\centering
\caption{Sensitivity to the ask cost $\kappa$ and the share of high-risk actions (independent world). Ask, WA, and U refer to the tuned fuzzy gate; the last column names the policy with the highest utility.}
\label{tab:sens}
\renewcommand{\arraystretch}{1.1}
\footnotesize
\begin{tabularx}{\columnwidth}{P{0.9cm} P{0.5cm} P{0.65cm} P{0.55cm} P{0.6cm} L}
\toprule
\textbf{High-risk} & $\boldsymbol{\kappa}$ & \textbf{Ask} & \textbf{WA} & \textbf{U} & \textbf{Best policy} \\
\midrule
40\% & 0.3 & 62.0 & 1.1 & 0.45 & Always confirm \\
40\% & 0.1 & 66.8 & 0.7 & 0.55 & Always confirm \\
40\% & 1.0 & 41.3 & 2.9 & 0.12 & Tuned fuzzy \\
20\% & 0.3 & 45.2 & 1.2 & 0.53 & Tuned threshold \\
10\% & 0.3 & 34.6 & 1.3 & 0.58 & Tuned threshold \\
\bottomrule
\end{tabularx}
\end{table}


\end{document}

%% file: preamble.tex
\def\instructgpt{\texttt{Instruct-GPT}\xspace}
\def\gptfour{\texttt{GPT-4}\xspace}
\def\gemini{\texttt{Gemini}\xspace}
\def\internlm{\texttt{InternLM2}\xspace}
\def\claude{\texttt{Claude 3}\xspace}
\def\reka{\texttt{Reka}\xspace}
\def\zephyr{\texttt{Zephyr}\xspace}
\def\deepseek{\texttt{DeepSeek-V2}\xspace}
\def\chatglm{\texttt{ChatGLM}\xspace}
\def\nemotron{\texttt{Nemotron-4 340B}\xspace}
\def\llama{\texttt{Llama 3}\xspace}
\def\qwen{\texttt{Qwen2}\xspace}
\def\gemma{\texttt{Gemma2}\xspace}
\def\starling{\texttt{Starling-7B}\xspace}
\def\athene{\texttt{Athene-70B}\xspace}
\def\hermes{\texttt{Hermes 3}\xspace}
\def\oone{\texttt{o1}\xspace}
\def\llms{\texttt{LLMs}\xspace}
\def\llm{\texttt{LLM}\xspace}
\def\rl{\texttt{RL}\xspace}
\def\rlhf{\texttt{RLHF}\xspace}
\def\rag{\texttt{RAG}\xspace}
\def\mdp{\texttt{MDP}\xspace}
\def\grpo{\texttt{GRPO}\xspace}
\def\trpo{\texttt{TRPO}\xspace}
\def\dpo{\texttt{DPO}\xspace}
\def\mle{\texttt{MLE}\xspace}
\def\rag{\texttt{RAG}\xspace}
\def\reinforce{\texttt{REINFORCE}\xspace}
\def\ppo{\texttt{PPO}\xspace}
\def\trpo{\texttt{TRPO}\xspace}
\def\mrt{\texttt{MRT}\xspace}
\def\seqgan{\texttt{SeqGAN}\xspace}
\def\mixer{\texttt{MIXER}\xspace}
\def\lora{\texttt{LoRA}\xspace}
\def\cot{\texttt{CoT}\xspace}
\def\rlaif{\texttt{RLAIF}\xspace}
\def\scst{\texttt{SCST}\xspace}

\newcommand{\insightbox}[1]{%
    \begin{tcolorbox}[colframe=black!70, colback=blue!5, boxrule=1pt, arc=4mm]
        \includegraphics[width=0.4cm]{figures/lightbulb.png}
        \textbf{\small#1}
    \end{tcolorbox}
}

\newcommand{\warningbox}[1]{%
    \begin{tcolorbox}[colframe=red!70, colback=yellow!10, boxrule=1pt, arc=4mm]
        \faExclamationTriangle\ \textbf{#1}
    \end{tcolorbox}
}